\documentclass[lettersize,journal]{IEEEtran}
\usepackage{amsmath,amsfonts}
\usepackage{algorithmic}
\usepackage{epsfig, algorithm}
\usepackage{array}
\usepackage{amssymb}
\usepackage{acronym} 
\usepackage[caption=false,font=normalsize,labelfont=sf,textfont=sf]{subfig}
\usepackage{textcomp}
\usepackage{stfloats}
\usepackage{tabularx}
\usepackage{xcolor}
\usepackage{url}
\usepackage{verbatim}
\usepackage{multirow}
\usepackage{acronym}
\usepackage{graphicx}	
\usepackage{stfloats}
\usepackage{soul}
\usepackage{cancel}
\usepackage{cite}
\def\BibTeX{{\rm B\kern-.05em{\sc i\kern-.025em b}\kern-.08em
T\kern-.1667em\lower.7ex\hbox{E}\kern-.125emX}}
\usepackage{balance}
\usepackage{multirow}
\usepackage{hhline}

\input{AcronymsListFinal}
\acresetall
\usepackage{fancyhdr}

\begin{document}
\title{Scalable Association of Users in CF-mMIMO: A Synergy of Communication, Sensing, and JCAS}
\author{Ahmed~Naeem, Anastassia~Gharib  \IEEEmembership{Member,~IEEE}, El Mehdi Amhoud   \IEEEmembership{Senior Member,~IEEE}, Hüseyin~Arslan~\IEEEmembership{Fellow,~IEEE}
\\This work has been submitted to the IEEE for possible publication. Copyright may be transferred without notice, after which this version may no longer be accessible.
}
\markboth{Journal of \LaTeX\ Class Files,~Vol.~14, No.~8, xxxx~20xx}%
{Shell \MakeLowercase{\textit{et al.}}: Bare Demo of IEEEtran.cls for IEEE Journals}
\maketitle 
\begin{abstract}
Cell-free massive multiple-input multiple-output (CF-mMIMO) is a key enabler for the sixth generation (6G) networks, offering unprecedented spectral efficiency and ubiquitous coverage. In CF-mMIMO systems, the association of user equipments (UEs) to access points (APs) is a critical challenge, as it directly impacts network scalability, interference management, and overall system performance. Conventional association methods primarily focus on optimizing communication performance. However, with the emergence of sensing and joint communication and sensing (JCAS) requirements, conventional approaches become insufficient. To address this challenge, we propose a scalable user association (SUA) scheme for CF-mMIMO networks, considering heterogeneous UE requirements. Designed to enhance the performance of both sensing and communication, the proposed SUA scheme aims to ensure network scalability. This is achieved by dynamically assigning APs to UEs based on their specific service requirements (communication, sensing, or JCAS), while considering link quality, interference mitigation, and network-related constraints. Specifically, the proposed SUA scheme employs AP masking, link prioritization, and an optimization-based association mechanism to select the most suitable APs for each UE. Simulations show that, compared to conventional CF-mMIMO methods, the proposed SUA scheme significantly reduces interference and computational runtime, while improving the symbol error rate for communication and the probability of detection for sensing.
\end{abstract}
\IEEEpeerreviewmaketitle
\section{Introduction}
\par As we progress towards the era of the \ac{6G} network, the demand for diverse applications and user requirements is rapidly increasing \cite{10559594}. The focus shifts toward the need to support an interconnected world where advanced applications like autonomous vehicles, smart cities, and industrial automation require not just communication but also precise sensing capabilities. \Ac{JCAS} exemplifies this by integrating communication and sensing within the same system. This seamless integration of sensing in \ac{6G} is essential for meeting the demands of future wireless networks, where communication and sensing work together to unlock new possibilities.
\par Meanwhile, \ac{CF-mMIMO} emerges as a key \ac{6G} enabler, providing a distributed network architecture capable of delivering consistent user-experienced data rates \cite{7827017} and high-precision sensing \cite{dhulashia2024multistatic} across diverse applications. Unlike traditional cellular systems, \ac{CF-mMIMO} eliminates the cell boundaries, where all \acp{AP} collaboratively serve \acp{UE}. This collaborative framework enhances \ac{SE}, making it suitable for ultra-dense \ac{6G} networks. Additionally, the distributed nature of \ac{CF-mMIMO} facilitates robust and accurate sensing capabilities by leveraging the multi-static radar principles \cite{dhulashia2024multistatic}, thereby addressing the dual requirements of communication and sensing integral to \ac{6G} networks.
\par To fully harness the benefits of \ac{CF-mMIMO} within the \ac{6G} framework, \ac{UA} is as a critical aspect. The initial concept of \ac{CF-mMIMO}, \cite{7827017, 7917284, 8845768, 8476516}, envisioned all \acp{AP} serving all \acp{UE}. While such an approach is theoretically promising, it is practically unscalable due to immense computational complexity, excessive front-haul load, and severe interference. Thus, in \ac{CF-mMIMO} systems, where multiple distributed \acp{AP} serve \acp{UE} without the constraints of traditional cell boundaries, efficient \ac{UA} is essential for optimizing network resources and ensuring seamless service delivery. To address this challenge, a \ac{UA} framework is required, wherein only the most suitable \acp{AP} serve each \ac{UE}, thereby enhancing system efficiency. Traditionally, \ac{UA} strategies focus on communication needs only, associating \acp{UE} to \acp{AP} based on communication metrics alone as in \cite{bjornson2020scalable, chen2020structured, liu2016joint}. While these approaches have proven to be effective for communication systems, they fall short of addressing the multifaceted requirements of \ac{6G}, including sensing and \ac{JCAS} capabilities. Recognizing this gap, our work introduces a novel \ac{SUA} scheme that integrates communication, sensing, and \ac{JCAS} needs, ensuring \acp{UE} are associated with the most optimal \ac{AP}(s) based on their specific requirements. This scalable approach reduces complexity and interference while meeting the diverse demands of \ac{6G} networks.
\subsection{Related Work}
\par This section reviews the related works on \ac{UA} in \ac{CF-mMIMO}. Traditional \ac{UA} strategies in \ac{CF-mMIMO} predominantly rely on \ac{LSF}-based methods~\cite{buzzi2017cell,rachuri2024novel} and competition-based schemes \cite{chen2020structured}, where \acp{UE} are associated with \acp{AP} based on channel conditions or by competing for network resources. For instance, the work in \cite{rachuri2024novel} proposes a \ac{UA} and power control scheme that dynamically assigns \acp{AP} to \acp{UE} based on \ac{LSF}, ensuring \ac{SE} requirements are met. Once \ac{UA} is established, a game-theoretic power control algorithm minimizes uplink power while satisfying \ac{SE} constraints. Similarly, the work in \cite{rachuri2024probabilistic} introduces a quality of service aware \ac{UA} scheme that leverages \ac{LSFC} to predict the minimum number of \acp{AP} required to meet user-specific \ac{SE} needs. Another approach presents a preference-based \ac{UA} method, where \acp{UE} associate with \acp{AP} offering the strongest \ac{LSFC}, while \acp{AP} prioritize \acp{UE} with higher \ac{LSFC} values~\cite{sarker2023access}. Additional works on \ac{LSF} \ac{UA} are discussed in \cite{le2021learning, chong2024performance}. Despite the significant improvements, these methods are limited to optimizing \ac{UA} based on communication needs, neglecting the emerging dual requirements of communication and/or sensing.
\par The integration of \ac{JCAS} in \ac{CF-mMIMO} systems for \ac{6G} networks demands a paradigm shift in \ac{UA} strategies. Recent works, such as \cite{demirhan2024cell, 10694569, buzzi2024scalability}, emphasized the importance of incorporating \ac{JCAS} capabilities. However, these studies primarily focus on beamforming, power allocation, and multi-target detection. Similarly, the work in \cite{mao2024communication} emphasizes defining the communication-sensing region to investigate the trade-off between these functionalities for beamforming strategies. Additionally, the work in \cite{behdad2022power} explores joint precoding and power allocation techniques. However, these studies largely overlook the necessity of adaptive \ac{UA} frameworks that cater to the demands of communication, sensing, or \ac{JCAS}, leaving a critical gap for comprehensive \ac{6G} system optimization. To address this gap, our work introduces a novel \ac{UA} framework tailored for \ac{CF-mMIMO} systems that simultaneously consider heterogeneous \ac{UE} requirements. By dynamically associating \acp{UE} with the most suitable \acp{AP} based on their specific service demands, the proposed \ac{SUA} scheme ensures optimized resource utilization, reduced latency, and improved performance for diverse applications integral to the \ac{6G}~vision.
\subsection{Paper Contributions and Organization}
This work proposes SUA, a scalable \ac{UA} scheme catering to the heterogeneous \ac{UE} requirements. Unlike conventional \ac{UA} approaches that prioritize only communication-related metrics, the proposed SUA scheme integrates diverse \ac{UE} service requirements 
to ensure a more comprehensive and adaptive UA~strategy. The contributions of this paper are as follows:
\begin{itemize}
\item To achieve optimal UA, the proposed SUA scheme first performs \ac{AP} masking, which eliminates weak and interference-prone links, reducing unnecessary connections and computational complexity while ensuring that only high-quality \ac{AP}-\ac{UE} associations are considered.
\item The proposed SUA scheme considers APs' preferences to be associated with specific UEs through link prioritization, where each AP strategically ranks the unmasked UEs based on their link quality and service demands to improve the communication and sensing~performance.
\item As part of the proposed SUA scheme, we then formulate a global optimization problem that maximizes weighted link quality while minimizing redundant connections. The optimization incorporates practical constraints including \ac{AP} capacity limits, \ac{UE} multi-AP association, AP masking, and prioritization, while ensuring feasibility.
\item We derive mathematical analysis of the  \ac{SER} and the probability of detection for the proposed \ac{SUA} scheme, which is then verified with simulations. The results demonstrate that, compared to the conventional unscalable \ac{CF-mMIMO} method, the proposed \ac{SUA} scheme reduces interference, enhancing \ac{SER} for communication and the probability of detection for sensing. 
\end{itemize}
\par The rest of the paper is organized as follows. Section~II discusses the system model, while Section~III introduces the proposed \ac{SUA} scheme. The performance evaluation analysis is discussed in Section~IV, and simulation results are presented in Section~V. Section~VI includes the conclusion. 

\begin{figure}
\centering 
\resizebox{0.95\columnwidth}{!}{
\includegraphics{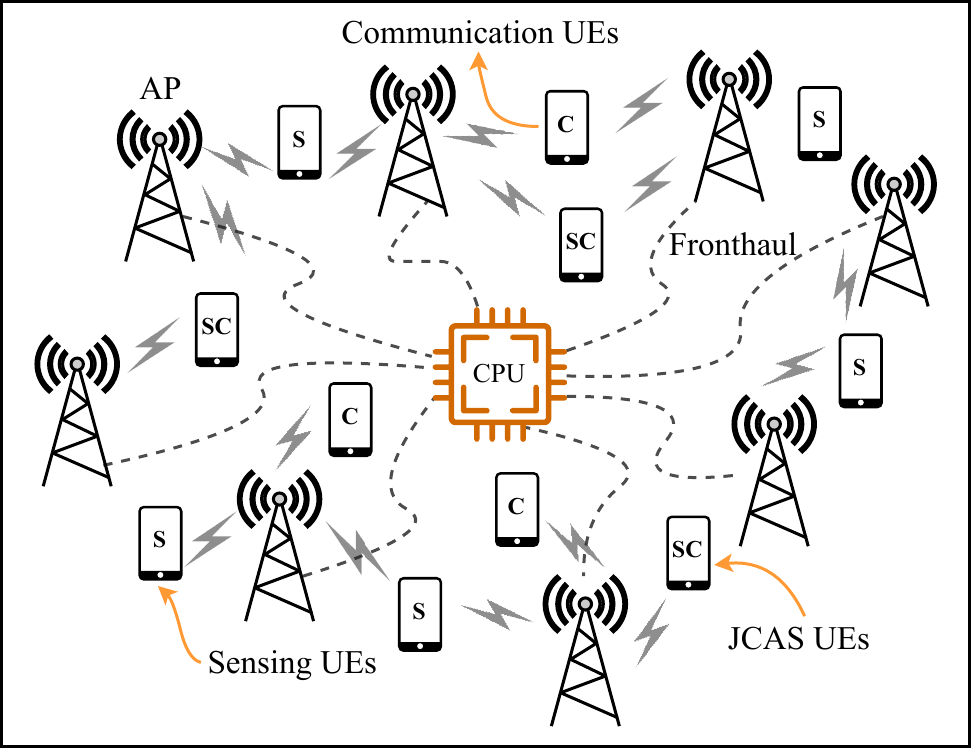}}
\caption{Proposed \ac{CF-mMIMO} system architecture.}
\label{FigMAIN1}
\end{figure}

\section{System Model}
\par This section presents the system model for the proposed \ac{SUA} scheme. The proposed \ac{CF-mMIMO} system architecture is shown in Fig. \ref{FigMAIN1}. It consists of $K$ \acp{UE} of heterogeneous requirements (based on the \ac{6G} vision) and  $L$ \acp{AP}, where~$L > K$. \acp{UE} include those with the requirements of communication (i.e., $\mu_k \in \mathcal{K}_{\text{com}}$), sensing (i.e., $\mu_k \in \mathcal{K}_{\text{sense}}$), and \ac{JCAS} (i.e., $\mu_k \in \mathcal{K}_{\text{JCAS}}$). 
The variable $\mu_k$ serves as a subset membership indicator, identifying whether a \ac{UE} $k$ belongs to the subset of UEs with communications $\mathcal{K}_{\text{comm}}$,  sensing $\mathcal{K}_{\text{sense}}$, or JCAS $\mathcal{K}_{\text{JCAS}}$ requirements.
Each \ac{UE} is equipped with a single antenna. Each AP is equipped with an array size of $N$ having $n$ antenna elements and is connected to a \ac{CPU} via a front-haul link~\cite{ajmal2024cell}. We assume that each \ac{AP} can serve a limited number of \acp{UE}, which depends on the available number of orthogonal pilots $\tau_p$. This limitation is essential to mitigate pilot contamination~\cite{kassam2025cell}. Additionally, we assume that every \ac{UE} is simultaneously served by $X$ \acp{AP}, where~$0<X<L$.  The limitation on $X$ prevents excessive \ac{AP}-\ac{UE} connections, reducing interference and complexity, while ensuring scalability. The system functions in \ac{TDD} mode, with coherent joint transmission and reception. The proposed \ac{UA} is to be performed collaboratively where both the \ac{AP} and \ac{CPU} take part in making the decisions on signaling and decoding. The propagation channels are modeled using a block-fading approach, where the channel remains constant within each coherence block of duration~$\tau_c$~\cite{demir2021foundations}. Next, we explain the signaling process for each \ac{UE}~requirement.
\subsubsection{Communication Signaling}
\par The channel response between the $l$-$th$ \ac{AP} and the $k$-$th$ \ac{UE}, $\mathbf{h}_{lk}\in\mathbb{C}^N$, is modeled as a random realization within each $\tau_c$, drawn from a stationary ergodic fading distribution. The $\mathbf{h}_{lk}$ follows a correlated Rayleigh fading distribution, $\mathbf{h}_{lk}\sim\mathcal{N}_\mathbb{C}(0,\mathbf{R}_{lk})$, with $\mathbf{R}_{lk}\in{\mathbb{C}}^{N\times{N}}$ representing the spatial correlation matrix, incorporating \ac{LSF}, \cite{demir2021foundations}. For \ac{UL} channel estimation, $\tau_p$ orthogonal pilot sequences are used, where $\tau_p$ is independent of $K$ \cite{bjornson2020scalable}. The received \ac{UL} pilot signal at the $l$-$th$ \ac{AP}~is: 
\begin{equation}
\mathbf{y}^{\mathrm{p}}_{lk}=\sum_{k=1}^K\sqrt{\tau_p p^p}~\mathbf h_{lk}+\mathbf{n}_{lk},
\end{equation}
where $p^p$ is the pilot transmit power, and $\mathbf{n}_{lk}$ is thermal noise. The \ac{MMSE} estimator provides that $\widehat{\mathbf{h}}_{lk}=\sqrt{p_k\tau_p}~\mathbf{R}_{lk}\Psi_{lk}^{-1}\mathbf{y}^{\mathrm{p}}_{lk}$, where $\Psi_{lk}$ is the correlation matrix of $\mathbf{y}^{\mathrm{p}}_{lk}$. While for \ac{UL} data, the received signal is:
\begin{equation}
\mathbf{y}^\mathrm{UL}_l=\sum_{k=1}^K\mathbf{h}_{lk}s_k+\mathbf{n}_l, 
\end{equation}
where $s_k$ is the transmitted signal from the $k$-$th$ \ac{UE}, and $\mathbf{n}_l\sim\mathcal{CN}(0,\sigma^2\mathbf{I}_N)$ is receiver noise with \ac{MR} combining applied at the $l$-$th$ \ac{AP}.
\begin{figure*}
\centering 
\resizebox{2\columnwidth}{!}{
\includegraphics{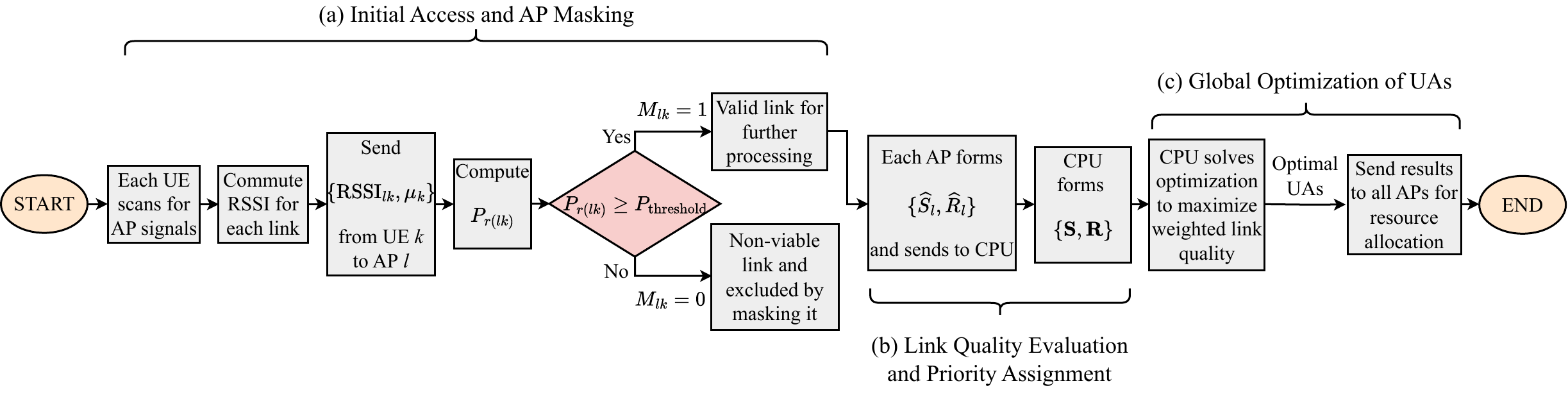}}
\caption{Detailed steps of the proposed \ac{SUA} scheme.}
\label{stepsabcd}
\end{figure*}
\subsubsection{Sensing Signaling}
\par In a monostatic radar, the target-free channel for the $l$-$th$ \ac{AP} is $\mathbf{H}_{l}$ with correlated Rayleigh fading. The random matrix $\mathbf{W}_{\mathrm{ch},l}$ contains independent and identically distributed (i.i.d.) entries following $\mathcal{CN}(0,1)$ distribution. Since the same \ac{AP} is used for transmission and reception, the spatial correlation matrix for the $l$-$th$ \ac{AP} is $\mathbf{R}_{l}$ and $\mathbf{H}_{l}=\mathbf{R}_{l}^{\frac12}\mathbf{W}_{\mathrm{ch},l}\mathbf{R}_{l}^{\frac12}$. The received signal at $m$ instance is:
\begin{equation} \label{mixone}
\begin{aligned}&\mathbf{y}_l[m]=\sum_{l=1}^{N}\underbrace{\alpha_{l}\sqrt{\beta_{l}}~\mathbf{a}(\phi_{0,l},\theta_{0,l})\mathbf{a}^T(\varphi_{0,l},\vartheta_{0,l})\mathbf{x}_l[m]}_{\text{desired reflections from the target}}\\&+\sum_{l=1}^{N}\underbrace{\mathbf{H}_{l}\mathbf{x}_l[m]}_{\text{clutter return}}+\mathbf{n}_l[m], ~~~\forall~~ m=1,..., L_d
\end{aligned}
\end{equation}
\noindent where $\mathbf{n}_l[m]$ is the receiver noise and the second term in Eq.~(\ref{mixone}) represents interference. The term $\beta_{l}$ includes the path-loss through the \ac{UE}, and $\sigma_\mathrm{rcs}$ - variance of the \ac{RCS}. Normalized \ac{RCS} follows Swerling-I model with $L_d$ sensing symbols. The array response vector is given by $\mathbf{a}(\varphi,\vartheta) = \begin{bmatrix}1  e^{j\pi\sin(\varphi)\cos(\vartheta)}  \ldots  e^{j(N-1)\pi\sin(\varphi)\cos(\vartheta)}\end{bmatrix}^T$, where $\varphi$ and $\vartheta$ are the azimuth and elevation angles, respectively.
\subsubsection{\ac{JCAS} Signaling}
\par When a \ac{UE} requires \ac{JCAS}, each \ac{AP} operates as a \ac{DFRC} system~\cite{liao2023robust}. Let $j(l)=[j_1(l),\ldots,j_G(l)]^{\mathsf{T}}$ denote the vector of $G$ data symbols for each \ac{UE}, with $\mathbb{E}[\mathbf{j}(l)\mathbf{j}^{\mathsf{H}}(l)] = \mathbf{I}_G$~\cite{liao2023robust}. Each \ac{AP} performs dual beamforming with $\mathbf{W}=[\mathbf{w}_1, \ldots, \mathbf{w}_G] \in \mathbb{C}^{N \times G}$, where the $k$-$th$ column $\mathbf{w}_k \in \mathbb{C}^N$ is the beamformer for \ac{UE} $k$. The transmitted signal from the \ac{DFRC} \ac{AP} is then:
\begin{equation}
\mathbf{x}(l)=\mathbf{W}\mathbf{j}(l)=\sum_{k=1}^K\mathbf{w}_kj_k(l),
\end{equation}
with the covariance matrix of $\mathbf{x}(l)$ given as $\mathbf{R} = \mathbb{E}[\mathbf{x}(l)\mathbf{x}^{\mathsf{H}}(l)] = \mathbf{W}\mathbf{W}^{\mathsf{H}} = \sum_{k=1}^{K}\mathbf{R}_k$, and the contribution of the $k$-$th$ beamformer given by $\mathbf{R}_k = \mathbf{w}_k\mathbf{w}_k^{\mathsf{H}}$.
\section{Proposed Scalable User Association Scheme}
\par This section outlines the proposed \ac{SUA} scheme presented in Fig. \ref{stepsabcd}. The process begins with each \ac{UE} accessing the network and finishes with each UE being associated with the optimal \ac{AP}(s). Specifically, the proposed \ac{SUA} scheme employs a structured approach starting with initial access and \ac{AP} masking in Fig.~\ref{stepsabcd}(a), where each \ac{UE} measures the \ac{RSSI} from the surrounding \acp{AP} and connects to \acp{AP} with the strongest channels.  With the help of the \ac{AP} masking step, each \ac{AP} ensures only links above a pre-defined received power threshold are considered. Following this, each \ac{AP} evaluates the link quality for the unmasked \acp{UE} depending on their service requirements. Based on the achieved values, every \ac{AP} assigns priorities for \ac{UA}s in  Fig.~\ref{stepsabcd}(b). This information is then sent by each \ac{AP} to the \ac{CPU} for global decision-making and optimization, finalizing the associations in Fig.~\ref{stepsabcd}(c). Such a hierarchical approach minimizes system complexity, improves resource allocation, and ensures diverse service requirements are optimally met. The following sections provide a detailed discussion of each step of the proposed \ac{SUA} scheme.
\begin{algorithm}[h]
\caption{Initial Access and AP Masking}\label{alg:initial_access_masking}
\begin{algorithmic}[1]
\STATE \textbf{Input:} $L$ \acp{AP}, $K$ \acp{UE}, $\mu_k \in \{\mathcal{K}_{\text{com}}, \mathcal{K}_{\text{sense}}, \mathcal{K}_{\text{JCAS}}\}~\forall~k = \{1, 2, \dots K\}$, 
$P_{\text{threshold}}$
\STATE \textbf{Output:} Masking vector $\widehat{M_l} \in \{0, 1\}^{K}$ $\forall$ AP $l=\{1, 2, \dots, L\}$ 
\vspace{1mm}
\STATE \textbf{Step 1: Initial Access}
\FOR{UE $k = 1$ to $K$} 
    \FOR{AP $l = 1$ to $L$} 
        \STATE Compute $\text{RSSI}_{lk}$ using Eq.~(\ref{plossfor})
        \STATE Send $\{\text{RSSI}_{lk}, \mu_k\}$ from UE $k$ to AP $l$ 
    \ENDFOR
\ENDFOR
\vspace{1mm}
\STATE Form $\widehat{\text{RSSI}}_l~\forall~\text{AP}~l,~\text{where}~l=\{1, 2, \dots,{L}\}$ 
\vspace{1mm}
\STATE \textbf{Step 2: AP Masking}
\FOR{AP $l = 1$ to $L$}
    \FOR{UE $k = 1$ to $K$}
         \STATE Compute $P_{r(lk)}$
        \IF{$P_{r(lk)} \geq P_{\text{threshold}}$}
            \STATE Set $M_{lk} \gets 1$
        \ELSE
            \STATE Set $M_{lk} \gets 0$
        \ENDIF
    \ENDFOR
    \STATE Form vector $\widehat{M}_l = \left[ M_{l1}, ~ M_{l2},  \ldots,  M_{lK} \right]\in \mathbb{R}^{K}$
\ENDFOR
\end{algorithmic}
\end{algorithm}
\subsection{Initial Access and \ac{AP} Masking}
\par Algorithm~\ref{alg:initial_access_masking} outlines the procedure for the initial access and \ac{AP} masking. The first step is the initial access, during which each \ac{UE} scans its surroundings to detect nearby \acp{AP}. This is done by receiving APs' periodically broadcasted synchronization signal blocks. Each \ac{UE} then  calculates the \ac{RSSI} between the detected \ac{AP}(s) and itself as follows~\cite{10643696}:
\begin{equation}\label{plossfor}
\text{RSSI}_{lk} =  P_t- \left( PL_0 + 10\gamma \log_{10}\left(\frac{d_{lk}}{d_0}\right)+ X_\delta \right),
\end{equation}
\noindent where $PL_0$ is the path loss at a reference distance $d_0$, $\gamma$~is the path loss exponent, $P_t$ is the transmit power of an \ac{AP}, $d_{lk}$ the distance between the $l$-$th$ \ac{AP} and the $k$-$th$ \ac{UE}, and $X_\delta$ represents fading effects modeled as a  Gaussian random variable with zero mean and unity standard deviation. Therefore, each $k$-$th$ \ac{UE} computes the \ac{RSSI} values for all detected \acp{AP} and generates a vector $\widehat{\text{RSSI}}_k = [\text{RSSI}_{1k}, \text{RSSI}_{2k}, \ldots, \text{RSSI}_{Lk}]^\top \in \mathbb{R}^L,$ which ranks the detected \acp{AP} based on link quality. These \ac{RSSI} measurements along with the service requirement indicators (i.e.,~$\{\text{RSSI}_{lk}, \mu_k\}$) are sent by each \emph{k-th} \ac{UE} to the corresponding \emph{l-th} \ac{AP}. From the received RSSI values, a joint vector, i.e.,  $\widehat{\text{RSSI}}_l = \left[ \text{RSSI}_{l1}, ~ \text{RSSI}_{l2},  \ldots,  \text{RSSI}_{lK} \right] \in \mathbb{R}^{K}$, is formed at each \ac{AP} to prepare for the \ac{AP} masking step.
\par Once the initial access is completed, the subsequent step is the \ac{AP} masking. This critical step allows \acp{AP} to potentially identify the best \ac{UE}(s) to serve based on a predefined threshold of acceptable received signal strength. Unlike unscalable \ac{CF-mMIMO} approaches \cite{7827017, 7917284, 8845768, 8476516}, the proposed \ac{AP} masking step acts as a pre-processing phase. It filters out links with poor channel conditions or high interference potential, significantly reducing the number of links. This approach not only reduces the interference and front-haul load but also simplifies the following optimization process by reducing the problem's dimensionality and enhancing the scalability and efficiency of the \ac{CF-mMIMO} network. The \ac{AP} masking decision is made by first computing the received signal power $P_{r(lk)}$ for each link between the $l$-$th$ \ac{AP} and the $k$-$th$ \ac{UE} in the same way as in \cite{9755972}. This value is then compared with a predefined threshold $P_\text{threshold}$ as follows:
\begin{equation}
M_{lk} =
\begin{cases}
1, & \text{if } P_{r(lk)} \geq P_\text{threshold}, \\
0, & \text{otherwise},
\end{cases}
\end{equation}
where $M_{lk}$ is the binary indicator that the $k$-$th$ \ac{UE} is masked by the $l$-$th$ \ac{AP} or not. $M_{lk} = 1$ signifies that the $P_{r(lk)}$ exceeds or equals the threshold $P_\text{threshold}$, making the link valid for further processing. Conversely, $M_{lk} = 0$ indicates that $P_{r(lk)}$ is below $P_\text{threshold}$, rendering the link non-viable and excluded by masking it. The predefined threshold value is typically set at -65 dBm \cite{miao2020estimating}. Since \acp{AP} perform the masking procedure with all initially connected \acp{UE} to them, each $l$-$th$ \ac{AP} then forms the $\text{M}_l$ vector with masking decisions of all its connected \acp{UE}, i.e.,  $\widehat{{M}}_l = \left[ M_{l1}, ~ M_{l2},  \ldots,  M_{lK} \right]\in \mathbb{R}^{K}$, where each element represents the masking decision for a specific \ac{UE}. Based on these decisions, each \ac{AP} proceeds with subsequent procedures to provide services exclusively to the masked \acp{UE}.
\subsection{Link Quality Evaluation and Priority Assignment}
\par 
The next phase in the proposed \ac{SUA} scheme involves each \ac{AP} computing the \ac{SNR} and \ac{SCNR} for the unmasked \acp{UE} (i.e., 
when $M_{lk}=1$). This is crucial for evaluating link quality and guiding the subsequent prioritization process. The \ac{SNR} serves as a key indicator of communication reliability and data transmission efficiency. While the \ac{SCNR} accounts for clutter interference, which is critical in sensing scenarios where environmental reflections can degrade detection accuracy. Each \ac{AP} identifies high-quality links by assessing these metrics to ensure robust service while avoiding poor-performing links. This step is indispensable for prioritizing links that help minimize \ac{SER} and maximize the sensing accuracy in the optimization phase. As shown in Algorithm~\ref{alg:snr_scnr_priority}, each \ac{AP} computes \ac{SNR}, \ac{SCNR}, or both, depending on the specific requirements of the \ac{UE} as~follows:
\begin{algorithm}[h]
\caption{Link Quality Evaluation and Priority Assignment}\label{alg:snr_scnr_priority}
\begin{algorithmic}[1]
\STATE \textbf{Input:} $K$ \acp{UE} $\in\{\mathcal{K}_{\text{com}}, \mathcal{K}_{\text{sense}}, \mathcal{K}_{\text{JCAS}}\}$,  $L$ \acp{AP}
\STATE \textbf{Output:} $\mathbf{S} \in \mathbb{R}^{L \times K}$ and $\mathbf{R} \in \mathbb{R}^{L \times K}$ at CPU
\vspace{1mm}
\STATE \textbf{Step 1: Link Quality Evaluation at \acp{AP}}
\FOR {AP $l = 1$ to $L$}
    \FOR{UE $k = 1$ to $K$}
        \IF {UE $k \in \mathcal{K}_{\text{com}}$ \& $M_{lk}=1$}
            \STATE Compute $\text{SNR}_{lk}$ using Eq. (\ref{algo11})
            \STATE Assign $S_{lk} \gets \text{SNR}_{lk}$
        \ELSIF{UE $k \in \mathcal{K}_{\text{sense}}$ \& $M_{lk}=1$}
            \STATE Compute $\text{SCNR}_{lk}$ using Eq. (\ref{algo12})
            \STATE Assign $S_{lk} \gets \text{SCNR}_{lk}$
        \ELSIF{UE $k \in \mathcal{K}_{\text{JCAS}}$ \& $M_{lk}=1$}
            \STATE Compute $\text{Joint}_{lk}$ using Eq. (\ref{algo13}) 
            \STATE Assign $S_{lk} \gets \text{Joint}_{lk}$
        \ELSIF{$M_{lk}=0$}
            \STATE Assign $S_{lk} \gets 0$
        \ENDIF
    \ENDFOR
    \STATE Form vector $\widehat{S}_{l}=[S_{l1}, ~ S_{l2},  \ldots, S_{lK}] \in \mathbb{R}^{K}$
\ENDFOR
\vspace{1mm}
\STATE \textbf{Step 2: Priority Assignments at \acp{AP}}
\FOR{AP $l = 1$ to $L$}
        \STATE Compute ${R}_{lk}~\forall~k \in K$ by normalizing $\widehat{S}_{l}$ across $k$
\STATE Form vector $\widehat{R}_{l}=[R_{l1}, ~ R_{l2},  \ldots, R_{lK}]\in \mathbb{R}^{K}$ at AP $l$
\STATE Send $\{\widehat{S}_l, \widehat{R}_l\}$ to CPU
\ENDFOR
\STATE The CPU forms matrix $\mathbf{S} \in \mathbb{R}^{L \times K}$ and $\mathbf{R} \in \mathbb{R}^{L \times K}$
\end{algorithmic}
\end{algorithm}
\subsubsection{Communication \acp{UE}} \ac{SNR} for each \ac{UE} $k \in \mathcal{K}_{\text{com}}$ is:
\begin{equation} \label{algo11}
\text{SNR}_{lk} = \frac{P_{r(lk)}}{N_0}, 
\end{equation} 
where $N_0$ represents the noise power.
\subsubsection{Sensing \acp{UE}} For every \ac{UE} $k \in \mathcal{K}_{\text{sense}}$, the \ac{SCNR} is:
\begin{equation} \label{algo12}
    \text{SCNR}_{lk} = \frac{P_{S(lk)}}{P_C + N_0},
\end{equation}
where $P_{S(lk)}$ is the power received from the $k$-$th$ \ac{UE} by the $l$-$th$ \ac{AP}, and $P_C$ is the power reflected from clutter~\cite{10824708}. 
\subsubsection{JCAS \acp{UE}}  When considering every \ac{UE} $k \in \mathcal{K}_{\text{JCAS}}$, it is crucial to balance the dual objective of \ac{JCAS}. Thus, a weighted combination of \ac{SNR} and \ac{SCNR} is utilized as follows:
\begin{equation} \label{algo13}
\text{Joint}_{lk} = w_c \cdot \text{SNR}_{lk} + w_s \cdot \text{SCNR}_{lk},
\end{equation}
where $w_c$ and $w_s$ are the weights assigned to the \ac{SNR} and \ac{SCNR}, respectively, based on the importance of communication and sensing in the application.
\par The metrics for all unmasked \acp{UE} are then added into a vector $\widehat{S}_{l}$ by each $l$-$th$ \ac{AP} as $\widehat{S}_{l}=[S_{l1}, ~ S_{l2},  \ldots, S_{lK}]$, where:
\begin{equation}
{S}_{lk} =
\begin{cases}
\text{SNR}_{lk}, & \text{if \ac{UE}}~ k \in \mathcal{K}_{\text{com}} ~\text{and}~M_{lk}=1, \\
\text{SCNR}_{lk}, & \text{if \ac{UE}}~ k \in \mathcal{K}_{\text{sense}}~\text{and}~M_{lk}=1, \\
\text{Joint}_{lk}, & \text{if \ac{UE}}~ k \in \mathcal{K}_{\text{JCAS}}~\text{and}~M_{lk}=1, \\
0, & \text{if}~M_{lk}=0.
\end{cases}
\end{equation}
Once each \ac{AP} forms its $\widehat{S}_{l}$ vector, the next step is to assign priorities to those \ac{UE}s based on their link quality. To facilitate efficient resource allocation, each \ac{AP} forms a priority vector $\widehat{R}_{l}$ by normalizing and sorting the entries of $\widehat{S}_{l}$ denoted as $\widehat{R}_{l}=[R_{l1}, ~ R_{l2},  \ldots, R_{lK}]\in \mathbb{R}^{K}$. Higher values in $\widehat{S}_{l}$ imply better link quality for the \ac{AP}-\ac{UE} pair and therefore, are given a high priority. The priority assigned by the $l$-$th$ \ac{AP} to the $k$-$th$ \ac{UE} is, $R_{lk} = \frac{S_{lk}}{\sum_{k \in K} S_{lk}}$, where ${R}_{lk} \in [0, 1]$ ensures that the priorities are normalized across all \ac{UE}s served by a specific $l$-$th$ \ac{AP}. Furthermore, after each $l$-$th$ \ac{AP} computes its $\widehat{S}_{l}$ and $\widehat{R}_{l}$, it sends these two inputs to the \ac{CPU}. The \ac{CPU} then integrates each \acp{AP} inputs to form two matrices; a) $\mathbf{R} = \left[ \widehat{{R}}_1; ~ \widehat{{R}}_2;  \ldots;  \widehat{{R}}_L \right] \in \mathbb{R}^{L \times K}$ and b) $\mathbf{S} = \left[ \widehat{{S}}_1; ~ \widehat{{S}}_2;  \ldots;  \widehat{{S}}_L \right] \in \mathbb{R}^{L \times K}$, where each column is the link quality value and priority for a single \ac{UE} across all $L$ \ac{AP}s, respectively. By leveraging these inputs, the global optimization process at the \ac{CPU} can focus on high-quality links, significantly enhancing overall network performance, while balancing the fairness of \ac{UA}s across all \acp{AP}.
\subsection{Global Optimization of UAs}
\par In the final step of the proposed \ac{SUA} scheme, the global optimization phase integrates all the computed metrics and priorities to determine UAs. 
The optimization aims 
to minimize interference by jointly maximizing link quality and APs' priorities subject to selecting a limited number of links considering the limitations of APs, UEs, and masking to improve the overall network performance and scalability. This optimization problem is formulated in the following way:
\begin{equation} \label{MAINOPTI}
\begin{aligned}
& \max_{a_{lk}} \sum_{l=1}^L \sum_{k=1}^K S_{lk} \cdot R_{lk} \cdot a_{lk} \\
\text{s.t.}~~~ & C1:~ \sum_{k=1}^K a_{lk} \leq \tau_p,~~~\forall~ \text{AP}~ l=\{1,2,...,L\},\\
& C2:~ \sum_{l=1}^L a_{lk} \leq X,~~~\forall~ \text{UE}~ k=\{1,2,...,K\},\\
& D3:~ a_{lk}=0~ \forall~ \{l, k\}, ~\text{where}~ M_{lk}=0. \\
\end{aligned}
\end{equation}
where the decision variable $a_{lk}$ is a binary decision variable that determines whether the $l$-$th$ \ac{AP} is selected to serve the $k$-$th$ \ac{UE} or not as follows:
\begin{equation}
a_{lk} =
\begin{cases}
1, & \text{if the} ~l\text{-}th~ \text{AP}~ \text{serves the}~k\text{-}th ~\text{UE}, \\
0, & \text{otherwise.}
\end{cases}
\end{equation}
In the objective function, the role of $S_{lk}$ is to represent the link quality metric, which quantifies the strength and reliability of the connection between the $l$-$th$ \ac{AP} and the $k$-$th$ \ac{UE}, incorporating factors such as \ac{SNR} and \ac{SCNR} depending of UE requirements. Meanwhile, $R_{lk}$ helps to account for the priority assigned by the $l$-$th$ \ac{AP} to the $k$-$th$ \ac{UE} in Algorithm~\ref{alg:snr_scnr_priority}. The optimization in Eq.~\eqref{MAINOPTI} is subject to multiple constraints and definitions. The constraints limit the solution space by imposing certain conditions on the decision variables (i.e., $a_{lk} \forall~l k$~pair). In contrast, definitions specify of set rules for these binary variables, making the problem more sparse and thus, less complex.  These constraints and definitions are:
\subsubsection{AP Capacity Constraints ($C1$)} Each $l$-$th$ \ac{AP} can serve a maximum of $\tau_p$ \acp{UE}. This ensures no pilot contamination occurs as the number of orthogonal pilots is limited.
\subsubsection{UE Multi-AP Association Constraint ($C2$)}  Each $k$-$th$ \ac{UE} can connect to a maximum of $X$ \acp{AP} simultaneously, allowing multi-AP association but preventing over-allocation of \ac{AP} resources to a single \ac{UE}. This constraint also helps with reducing the load over each \ac{UE} and improving scalability.
\subsubsection{Masking Definition ($D3$)} Only valid links defined by the \ac{AP} masking matrix $\mathbf{M}$ are considered in the optimization. If $M_{lk} = 1$, the corresponding decision variable $a_{lk}$ is considered and can take the value of $0$ or $1$, indicating an inactive~or active link, respectively. Conversely, if $M_{lk} = 0$, decision variable $a_{lk}$ is forced to $0$, effectively masking the link and excluding it from further consideration in the optimization. This is because the link does not meet the required signal quality threshold, making it unsuitable for link establishment and ensuring that optimization focuses on high-quality links.
\par The optimization in Eq. (\ref{MAINOPTI}) is a mixed-integer linear programming problem with linear constraints and binary variables \cite{gharib2020scalable}. To solve this problem, the branch-and-bound technique is used to search the complete solutions’ space for the optimal solution. The solution of the optimization is an optimal matrix $\mathbf{A}=\{a_{lk}\} \in \mathbb{R}^{L \times K}$ that specifies the association decisions. These decisions are sent back to the respective \acp{AP} by the \ac{CPU}, which then configure resources accordingly. By integrating the computed metrics, priorities, and constraints, this optimization ensures a scalable, efficient, and high-performance \ac{CF-mMIMO} network tailored for diverse requirements of \acp{UE}.
\par The feasibility of the optimization problem in Eq. (\ref{MAINOPTI}) is ensured through constraints ($C1$, $C2$) and definition ($D3$), which align with realistic \ac{CF-mMIMO} system limitations, such as finite pilots, a manageable number of \acp{AP} per \ac{UE}, and masking based on link quality, respectively. The constraints ($C1$, $C2$) allocate resources to feasible \ac{AP}-\ac{UE} links by specifying an upper-bound for the number of links to be selected, avoiding over-allocation, interference, and inefficiencies. The masking definitions ($D3$) eliminates weak, cluttered links, enhancing performance. This ensures the solution maximizes weighted link quality while remaining feasible and practical for real-world scenarios, demonstrating the scalability and adaptability of the proposed \ac{SUA} scheme in dense networks.
\subsection{Complexity Analysis of \ac{SUA}}
\par In this section, we derive the complexity for the proposed \ac{SUA} scheme and compare it with the unscalable \ac{CF-mMIMO} of~\cite{7827017, 7917284, 8845768, 8476516}. The analysis focuses on three components: link quality assessment, \ac{UA} decisions, and resource allocation.
\subsubsection{Complexity of Link Quality Assessment}
For each \ac{AP}-\ac{UE} pair, link quality is normally evaluated using metrics like \ac{RSSI}, \ac{SNR}, or \ac{SCNR}. With $L \times K$ total pairs, the path loss and \ac{SNR}/\ac{SCNR} for each link is computed with $O(1)$ complexity per link. Thus, the total complexity in the conventional unscalable \ac{CF-mMIMO} is $O(L \times K)$. In the proposed \ac{SUA} scheme, \ac{AP} masking reduces the number of considered links to a fraction of $\psi$ (with $\psi \ll 1$), where $\psi$ depends on network conditions. It reflects the proportion of viable links, which decreases in denser networks with more scatterers and interference. This reduces the effective complexity to $O(\psi . (L \times K))$. Thus, the sparsity introduced by \ac{AP} masking minimizes the input size for optimization, lowering the computational load. In contrast, unscalable \ac{CF-mMIMO} systems evaluate all $L \times K$ links without masking, maintaining a complexity of $O(L \times K)$, thereby increasing the computational burden on \acp{AP}.
\subsubsection{Complexity of \ac{UA} Decisions}
The proposed \ac{SUA} scheme uses the optimization in Eq.~(\ref{MAINOPTI}) to perform UA.  Eq.~(\ref{MAINOPTI}) involves $L \times K$ binary variables and $L + K$ constraints. The definitions in Eq.~(\ref{MAINOPTI})-D3 reduces the number of decision variables to $\psi \cdot (L \times K)$. Thus, the optimization has a complexity of $O(\psi \cdot (L \times K) + L + K)$. For large $L$ and $K$,   $\psi$ is smaller due to excessive interference. Thus, the complexity can be approximated to $O(L + K)$ in this case. While in unscalable \ac{CF-mMIMO} systems, all $L \times K$ links are considered without masking, requiring no optimization. Since the complexity per link is $O(1)$, the complexity for \ac{UA} decisions is $O(L \times K)$.
\subsubsection{Complexity of Resource Allocation}
After \ac{UA}s are determined, resources (e.g., power, bandwidth) are allocated at each \ac{AP}. For the proposed SUA scheme, each \ac{UE} is served by up to $X$ \acp{AP}, so the number of active links is at most $X \times K$. This results in a complexity of approximately $O(X \times K)$. However, in conventional unscalable \ac{CF-mMIMO} systems, the complexity is $O(L \times K)$. It is worth noting that $L>>X$. Thus, \ac{SUA} has lower resource allocation complexity.
\subsubsection{Total Complexity}
The proposed \ac{SUA} scheme achieves a total complexity of $O(2\cdot \psi \cdot (L \times K)) + O(L + K) + O(X \times K)$, which is lower than the fixed $3 \cdot O(L \times K)$ of the conventional unscalable \ac{CF-mMIMO}. 
This results in computational savings, especially for systems with many \acp{AP} and \acp{UE} (i.e., large $L$ and $K$ value leading to smaller $\psi$), ensuring scalability.
\section{Mathematical Performance Analysis of the Proposed Scalable User Association Scheme}
\par In this section, performance  analysis of the proposed \ac{SUA} scheme is compared with the existing unscalable \ac{CF-mMIMO} systems \cite{7827017, 7917284, 8845768, 8476516}   mathematically. For this purpose, we evaluate the \ac{SER} and \ac{$P_d$} of the proposed SUA scheme, demonstrating enhanced performance and accuracy for communication and sensing, irrespectively of UE requirements (sensing, communication, or JCAS).
\subsection{Analysis of Symbol Error Rate}
\par This section presents a detailed \ac{SER} derivation  for both the proposed \ac{SUA} scheme and unscalable \ac{CF-mMIMO} systems. A key challenge in \ac{SER} is the non-identical \ac{LSF}~and \ac{CEE}, leading to a complex \ac{cPEP}. In unscalable \ac{CF-mMIMO}, higher \ac{LSF} and \ac{CEE} variations degrade performance. In this section we will show that the proposed \ac{SUA} scheme mitigates this by selecting fewer but stronger links, reducing these effects.
\par In the proposed \ac{SUA} scheme, links are selected based~on high \ac{SNR}, while in the unscalable systems, all the links~contribute, including those with poor \ac{SNR} amplifying the effects of non-identical \ac{LSF} and \ac{CEE}. To compute the \ac{unPEP}, we must average the \ac{cPEP} over all possible channel realizations. However, the \ac{PDF} of the channel-related variables especially in the unscalable system is difficult to determine due to the sheer number of weak links contributing to interference. In the proposed \ac{SUA} scheme, the exclusion of weak links results in a more manageable \ac{PDF}, simplifying the derivation process.
\subsubsection{Conditioned Pairwise Error Probability}
\par The goal is first to derive the \ac{cPEP} for a \ac{CF-mMIMO} system where a \ac{ML} detector is employed. Once the detection is completed, this is called \ac{cPEP}. The following theorem derives the \ac{cPEP} for the proposed \ac{SUA} scheme.
\par \textbf{Theorem 1.} \textit{For the proposed \ac{SUA} scheme with any modulation, the \ac{PEP} upper bound of \ac{ML} detector based on channel estimation is denoted as $\mathrm{Pr}(s_k^i\rightarrow s_k^j \mid \mathbf{\hat{h}}_{lk})$ and defined as:}
\begin{equation} \label{pp}
Q\left(
\frac{\|\mathbf{\hat{h}}_{lk}(s_k^i - s_k^j)\|^2}
{\sqrt{2 \left[\mathbf{\hat{h}}_{lk}(s_k^i - s_k^j)\right]^H \Sigma \left[\mathbf{\hat{h}}_{lk}(s_k^i - s_k^j)\right]}}
\right).
\end{equation}
(\textit{Proof.} See Appendix).
\par For simplicity, we derive an asymptotic bound of Eq. (\ref{pp}). The covariance matrix of the error $\mathbf{B}_k$ depends on path loss, noise variance, allocated power, $\tau_p$, and $X$. In the proposed \ac{SUA} scheme, only high-quality \ac{AP}-\ac{UE} links are selected with high \ac{SNR}, in this case, $\mathbf{B}_k$ for the $l$-$th$ \ac{AP} and the $k$-$th$ \ac{UE} is $\mathbf{B}_{lk}= \frac{\sigma^2 X}{\tau_p p_k} \mathbf{I}_N$. By focusing only on high-SNR links and limiting the $X$ and $\tau_p$, the proposed \ac{SUA} scheme significantly reduces the impact of channel estimation errors. This approach contrasts the unscalable system, which includes links with poor \ac{SNR}, thereby having higher estimation errors. Furthermore, the total covariance matrix $\Sigma$ accounts for, the noise variance ($\sigma^2$), and the accumulated effects of $\mathbf{B}_{k}$ across all \acp{UE}. For the proposed \ac{SUA} scheme only high-quality links are considered, making $\sum_{k=1}^K p_k \mathbf{B}_{lk}$ sparse. Substituting $\mathbf{B}_{lk}$ into $\Sigma$ as:

\begin{equation} 
\Sigma = \sigma^2 \mathbf{I}_{N} + \sum_{k=1}^K p_k \frac{\sigma^2 X}{\tau_p p_k} \mathbf{I}_N =\left(\sigma^2 + \frac{\sigma^2 X K}{\tau_p}\right) \mathbf{I}_{N},
\end{equation}
\noindent this indicates that $\Sigma$ is diagonal and its values depend on $K$, $X$ and $\tau_p$. Furthermore, Eq. (\ref{pp}) can be written as:
\begin{equation} 
\Pr\left(s_k^i \to s_k^j \big| \mathbf{\hat{h}}_{lk}\right) = Q\left(
\frac{\|\mathbf{\hat{h}}_{lk}(s_k^i - s_k^j)\|^2}{2(\sigma^2 + b^2)}
\right),
\end{equation}
\noindent where $b^2 = \frac{\sigma^2 X K}{\tau_p}$. The proposed \ac{SUA} scheme ensures stronger links ($\mathbf{\hat{h}}_{lk}$), increasing \(\|\mathbf{\hat{h}}_{lk}(s_k^i - s_k^j)\|^2\). Additionally, by limiting $X$, $\tau_p$, and applying \ac{AP} masking, the $\Sigma$ becomes sparser, effectively reducing the impact of interference.
\subsubsection{Unconditioned Pairwise Error Probability}
By averaging the \ac{cPEP} over the channel realizations we get the \ac{unPEP}:
\begin{equation} \label{ssk}
\Pr(s_k^i \to s_k^j) = \int_{0}^{\infty} Q\left(\frac{\gamma}{\sqrt{2D}}\right) f_\gamma(\gamma) \, d\gamma,
\end{equation}
where, $\gamma = \|\mathbf{\hat{h}}_{lk}(s_k^i - s_k^j)\|^2$, $f_\gamma(\gamma)$ is the PDF of $\gamma$, and $D = \left[\mathbf{\hat{h}}_{lk}(s_k^i - s_k^j)\right]^H \Sigma \left[\mathbf{\hat{h}}_{lk}(s_k^i - s_k^j)\right]$. Now by approximating $Q(x) \approx \frac{1}{12}e^{-x^2/2} + \frac{1}{4}e^{-2x^2/3}$ \cite{xiao2020generalized}, rewrite Eq. (\ref{ssk})~as:
{\small
\begin{equation}
\begin{aligned}
\Pr(s_k^i \to s_k^j) &\approx \int_{0}^{\infty} \left[\frac{1}{12}e^{-\frac{\gamma^2}{4D}} + \frac{1}{4}e^{-\frac{\gamma^2}{3D}}\right] f_\gamma(\gamma) \, d\gamma \\
&\approx \frac{1}{12} \int_{0}^{\infty} e^{-\frac{\gamma^2}{4D}} f_\gamma(\gamma) \, d\gamma
+ \frac{1}{4} \int_{0}^{\infty} e^{-\frac{\gamma^2}{3D}} f_\gamma(\gamma) \, d\gamma.
\end{aligned}
\end{equation}
}

\par For any modulation, $s_k^i - s_k^j = \Delta^{ij} = [\Delta^{ij}_k]_{k=1}^K$, where $\Delta^{ij}_k$ is the symbol difference. The distribution of $\gamma$ is influenced by $\Delta^{ij}$ as it dictates the contribution of the \acp{UE} to the effective signal strength, $\gamma$ is represented as $\gamma = \left| \sum_{l=1}^L \sum_{n=1}^N \sum_{k=1}^K h_{l,n,k} \Delta^{ij}_k \right|^2$, with effective variance of $\alpha_{lk} = \frac{p_k \tau_p \beta^2_{lk}}{p_k \tau_p \beta_{lk} +X \sigma^2}$, $f_\gamma(\gamma) \sim \text{sum of exponential variables}$. simplify $f_\gamma(\gamma)$, we use~\ac{MGF}:
\begin{equation}
M_\gamma(t) = \prod_{l=1}^L \prod_{n=1}^N \left(1 - t \sum_{k=1}^K \alpha_{lk} |\Delta^{ij}_k|^2 \right)^{-1}.
\end{equation}
\noindent Substituting $M_\gamma(t)$ into the integral for the \ac{unPEP}:
\begin{equation}
\Pr(s_k^i \to s_k^j) \approx \frac{1}{12} M_\gamma\left(-\frac{1}{4D}\right) + \frac{1}{4} M_\gamma\left(-\frac{1}{3D}\right).
\end{equation}
\par In the proposed \ac{SUA} scheme, the sparse association reduces the summation in $\alpha_{lk}$, making:
\begin{equation} \label{Mimi}
M_\gamma(t) = \prod_{l=1}^L \left( 1 - t \sum_{k \in \mathcal{K}} \alpha_{lk} |\Delta^{ij}_k|^2 \right)^{-N},
\end{equation}
where $\mathcal{K}$ is the set of associated \acp{AP} for the proposed scheme (sparse association). Thus, the \ac{unPEP} becomes:
\begin{equation}
\begin{aligned}
\Pr(s_k^i \to s_k^j) = & \frac{1}{12} \prod_{l=1}^L \left( 1 + \frac{1}{4D} \sum_{k \in \mathcal{K}} \alpha_{lk} |\Delta^{ij}_k|^2 \right)^{-N}
+\\& \frac{1}{4} \prod_{l=1}^L \left( 1 + \frac{1}{3D} \sum_{k \in \mathcal{K}} \alpha_{lk} |\Delta^{ij}_k|^2 \right)^{-N}.
\end{aligned}
\end{equation}
The final \ac{unPEP} expression for the proposed \ac{SUA} scheme is:
{\small
\begin{equation} \label{UNCON}
\begin{aligned}
&\Pr(s_k^i \to s_k^j) = \frac{1}{12} \prod_{l=1}^L \left( 1 + \frac{1}{4(2(\sigma^2 + c^2))} \sum_{k \in \mathcal{K}} \alpha_{lk} |\Delta^{ij}_k|^2 \right)^{-N} \\&
+ \frac{1}{4} \prod_{l=1}^L \left( 1 + \frac{1}{3(2(\sigma^2 + c^2))} \sum_{k \in \mathcal{K}} \alpha_{lk} |\Delta^{ij}_k|^2 \right)^{-N},
\end{aligned}
\end{equation}
}
\noindent where \(c^2 = \frac{\sigma^2 K}{\tau_p X}\). This derivation considers sparse association and high-\ac{SNR} conditions to simplify $\alpha_{lk}$, leveraging the proposed scheme's optimization to suppress weak links.
\subsubsection{Symbol Error Rate}
\par The \ac{SER} is obtained by combining the \ac{unPEP} over all possible transmitted symbols. The \ac{SER} for M-ary modulation schemes can be expressed in terms of the average \ac{PEP}, which is the probability that a symbol $s_k^i$ is mistaken for another symbol $s_k^j$, which can be expressed as:
\begin{equation} \label{ser_general}
{\text{SER}}=\frac{1}{M} \sum_{i=1}^M \sum_{\substack{j=1 \\ j \neq i}}^M \Pr(s_k^i \to s_k^j),
\end{equation}
where $M$ is the constellation points, $s_k^i$ and $s_k^j$ represents transmitted symbols, and $\Pr(s_k^i \to s_k^j)$ is the \ac{unPEP} derived in Eq. (\ref{UNCON}). Substituting Eq. (\ref{UNCON}) into Eq. (\ref{ser_general}) becomes:
\begin{equation} \label{mi2}
\text{SER} = \frac{1}{M} \sum_{i=1}^{M} \sum_{\substack{j=1 \\ j \neq i}}^{M} 
\left[ \frac{1}{12} M_{\gamma}\left(-\frac{1}{4D}\right) 
+ \frac{1}{4} M_{\gamma}\left(-\frac{1}{3D}\right) \right],
\end{equation}
substituting Eq. (\ref{Mimi}) into Eq. (\ref{mi2}) we get final \ac{SER}~expression:
{\small
\begin{equation} \label{mainSERR}
\begin{aligned}
\text{SER} &= \frac{1}{M} \sum_{i=1}^{M} \sum_{\substack{j=1 \\ j \neq i}}^{M} 
\left[ \frac{1}{12} \prod_{l=1}^{L} \left( 1 + \frac{\sum_{k \in \mathcal{K}} \alpha_{lk} |\Delta_{k}^{ij}|^2}{4D} \right)^{-N} \right. \\ 
& \quad + \left. \frac{1}{4} \prod_{l=1}^{L} \left( 1 + \frac{\sum_{k \in \mathcal{K}} \alpha_{lk} |\Delta_{k}^{ij}|^2}{3D} \right)^{-N} \right] \\[5pt]
&= \frac{1}{12M} \sum_{i=1}^{M} \sum_{\substack{j=1 \\ j \neq i}}^{M} \prod_{l=1}^{L} 
\left( 1 + \frac{\sum_{k \in \mathcal{K}} \alpha_{lk} |\Delta_{k}^{ij}|^2}{4 \big(2(\sigma^2 + c^2)\big)} \right)^{-N} \\ 
& \quad + \frac{1}{4M} \sum_{i=1}^{M} \sum_{\substack{j=1 \\ j \neq i}}^{M} \prod_{l=1}^{L} 
\left( 1 + \frac{\sum_{k \in \mathcal{K}} \alpha_{lk} |\Delta_{k}^{ij}|^2}{3 \big(2(\sigma^2 + c^2)\big)} \right)^{-N}.
\end{aligned}
\end{equation}}
\subsection{Probability of Detection}
\par In this section, we derive \ac{$P_d$}. Once the matched filtering is performed over each delay and Doppler bin, of particular interest is the \ac{$P_d$} of the delay-Doppler bin where the \ac{UE} lies. To model this, we formulate the \ac{$P_d$} problem at the $l$-$th$ \ac{AP} as a binary hypothesis-testing problem \cite{hariri2015joint}:
\begin{equation} \label{PAH}
\begin{cases}
H_0:y_l=\Phi(c_l+n_l)~~~~~~~~~~~~~(\textrm{Target Absent})      \\H_1:y_l=\Phi(\sigma_t s+c_l+n_l)~~~~~(\textrm{Target Present})&
\end{cases},
\end{equation}
where $y_l$ is the measurement vector, $\Phi$ is a real random Gaussian matrix with i.i.d. elements, $c_l$ is clutter, and noise term as $n_l$. The Rayleigh-distributed \ac{RCS} of the \ac{UE} with the signal $s$ having a delay $\tau$ and Doppler shift $f_d$ is $\sigma_t \sim \mathcal{CN}(0, 1)$.
\par In the case if the \ac{UE} exists $\tau, f_d$ are not known \textit{a priori}, so the usual likelihood ratio test cannot be computed for the detection, instead the \ac{GLRT} is used \cite{5393299}. The likelihood functions under $H_0$ and $H_1$ are:
\begin{equation}
\begin{aligned}
&f(y_l|H_0) = \frac{1}{\pi^M |\Sigma_0|} \exp\left(-y^\text{H} \Sigma_0^{-1} y\right), \\
&f(y_l|H_1) = \frac{1}{\pi^M |\Sigma_1|} \exp\left(- (y - \Phi s)^\text{H} \Sigma_1^{-1} (y - \Phi s) \right),
\end{aligned}
\end{equation}
where the covariance matrix under $H_0$ and $H_1$ is $\Sigma_0 = \sigma_c^2 \mathbf{I} + \sigma_n^2 \mathbf{I}$ and $\Sigma_1 = \sigma_c^2 \mathbf{I} + \sigma_n^2 \mathbf{I}$, respectively. The \ac{GLRT} tests the ratio of the two likelihood functions as:
\begin{equation} \label{GLRT}
\Lambda(y_l) = \frac{f(y_l|H_1)}{f(y_l|H_0)} \underset{H_0}{\overset{H_1}{\gtrless}} \eta
\end{equation}
where $\eta$ is the detection threshold. Simplifying Eq. (\ref{GLRT}) as:
\begin{equation} \label{exp}
\ln \Lambda(y_l) = -y^\text{H} \Sigma_0^{-1} y + (y - {\Phi} s)^\text{H} \Sigma_1^{-1} (y - {\Phi} s),
\end{equation}
since $\Sigma_0$ and $\Sigma_1$ are identical Eq. (\ref{exp}) becomes as, $\ln \Lambda(y_l) = -y_l^\text{H} \Sigma^{-1} y_l+ (y_l - {\Phi} s)^\text{H} \Sigma^{-1} (y_l - {\Phi} s) = -y_l^\text{H} \Sigma^{-1} y_l + \big[ y_l^\text{H} \Sigma^{-1} y_l - 2 \operatorname{Re} \big\{ y_l^\text{H} \Sigma^{-1} {\Phi} s \big\} + ({\Phi} s)^\text{H} \Sigma^{-1} ({\Phi} s) \big]= -2 \operatorname{Re} \big\{ y_l^\text{H} \Sigma^{-1} {\Phi} s \big\} + ({\Phi} s)^\text{H} \Sigma^{-1} ({\Phi} s)$. Finally we get, $\ln \Lambda(y_l) = 2 \operatorname{Re} \left\{ y_l^\text{H} \Sigma^{-1} {\Phi} \mathbf{s} \right\} - ({\Phi} s)^\text{H} \Sigma^{-1} ({\Phi} s)$, thus:
\begin{equation} \label{12}
\Lambda(y_l) = 2 \operatorname{Re} \left\{ y_l^\text{H} \Sigma^{-1} {\Phi} s \right\} \underset{H_0}{\overset{H_1}{\gtrless}} \eta.
\end{equation}
\par In the proposed \ac{SUA} scheme, multiple \acp{AP} sense the same target. The combined test statistic across all the \acp{AP} in $\mathcal{L}_k$ is:
\begin{equation} \label{123}
\Lambda = \sum_{l \in \mathcal{L}_k} \Lambda(y_l),
\end{equation}
Further by substituting Eq. (\ref{12}) into Eq. (\ref{123}) we have, $\Lambda = \sum_{l \in \mathcal{L}_k} \left[ 2 \operatorname{Re} \left\{ y_l^\text{H} \Sigma^{-1} {\Phi} s \right\} - ({\Phi} s)^\text{H} \Sigma^{-1} {\Phi} s \right]$. In the case of the unscalable \ac{CF-mMIMO}, the decision rule becomes $\Lambda \underset{H_0}{\overset{H_1}{\gtrless}} \eta.$ Under $H_0$, $z = |y_l|$ follows a Rayleigh distribution because the signal contains only clutter and noise. The PDF of $z$ is:
\begin{equation}
p_z(z|H_0) =
\begin{cases}
\frac{2z}{\sigma_{\varphi}^2} \exp\left( -\frac{z^2}{\sigma_{\varphi}^2} \right), & z \geq 0, \\
0, & z < 0,
\end{cases}
\end{equation}
where $\sigma_{\varphi}^2 = \sigma_c^2 + \sigma_n^2$ represents the total power of clutter and noise. The probability of a false alarm is the probability that $\Lambda$ exceeds a detection threshold $\eta$ under $H_0$, $P_{FA} = \mathbb{P}(\Lambda > \eta| H_0)$. By integrating the Rayleigh \ac{PDF} from $\eta$ to $\infty$, we get $P_{FA} = \exp\left( -\frac{\eta^2}{\sigma_{\varphi}^2} \right)$. Thus, $\eta = \sigma_{\varphi} \sqrt{-\ln P_{FA}}.$
\par As the detection decision is made for each $\tau, f_d$ bin, $\eta$ can be adjusted accordingly based on $\sigma_{\varphi}$ to maintain a constant $P_{FA}$. While when a \ac{UE} exists under $H_1$, $z = |y_l|$ follows Rician distribution due to the presence of the target signal:
\begin{equation}
p_z(z|H_1) =
\begin{cases}
\frac{2z}{\sigma_{\varphi}^2} \exp\left( -\frac{z^2 + m^2}{\sigma_{\varphi}^2} \right) I_0 \left( \frac{2mz}{\sigma_{\varphi}^2} \right), & z \geq 0, \\
0, & z < 0,
\end{cases}
\end{equation}
where $m = |\sigma_t| \sqrt{P_S}$ is the non-centrality parameter, and $I_0(\cdot)$ is the modified Bessel function. The \ac{$P_d$} that $\Lambda$ exceeds the $\eta$ under $H_1$, $P_d = \mathbb{P}(\Lambda > 2\eta | H_1)$. Using the Rician distribution, \ac{$P_d$} is expressed in terms of the Marcum Q-function:
\begin{equation} \label{pd_combined}
\begin{aligned}
P_d &= Q_1\left( \sqrt{\frac{2m^2}{\sigma_{\varphi}^2}}, \sqrt{\frac{2\eta^2}{\sigma_{\varphi}^2}} \right), \\[3.5pt]
    &= Q_1\left( \sqrt{\frac{2(|\sigma_t|^2 \sqrt{P_S})}{\sigma_{\varphi}^2}}, \sqrt{\frac{2(\sigma_{\varphi} \sqrt{-\ln P_{FA}})^2}{\sigma_{\varphi}^2}} \right), \\[3.5pt]
    &= Q_1\left( \sqrt{2 \cdot \text{SCNR}_k}, \sqrt{-2 \ln P_{FA}} \right),
\end{aligned}
\end{equation}
where the \ac{$P_d$} increases with the output \ac{SCNR}. To derive the \ac{$P_d$} for distributed \acp{AP}, we extend Eq. (\ref{pd_combined}) by incorporating the contributions from all \acp{AP}. The aggregate \ac{$P_d$} is expressed as, $P_d = Q_1\left(\sqrt{2 \cdot \sum_{l=1}^{L} \text{SCNR}_l}, \sqrt{-2 \ln P_{FA}}\right)$. Higher aggregate \ac{SCNR} enhances the system’s ability to detect the target reflecting the overall sensing performance of the network.
\section{Simulation Results}
\par In this section, simulation results of the proposed \ac{SUA} scheme are presented from a network perspective. A simulation setup with $L=100$ \acp{AP} and $K=30$ \acp{UE} are considered, which are independently and uniformly distributed in a coverage area of 0.5 km $\times$ 0.5 km square in a random manner. The \acp{UE} are with diverse service requirements, including $24\%$ communication \acp{UE}, $40\%$ sensing \acp{UE}, and $36\%$ \ac{JCAS} \acp{UE}. For \acp{UE} with \ac{JCAS} requirement, we assume $w_c=0.4$ and $w_s=0.6$. Higher weight for sensing is set as sensing is more sensitive to \ac{AP}-\ac{UE} associations, while communication can adapt through modulation and coding and still have reliable transmissions \cite{10559594}. All \acp{AP} are equipped with half-wavelength-spaced uniform linear arrays, each of $N=5$ antennas. All \acp{AP} are also connected via a front-haul to a \ac{CPU}. We assume a distributed implementation, following \ac{TDD} protocols as in~\cite{bjornson2020scalable}. To compute the large-scale propagation conditions, such as pathloss and shadow fading, the 
urban microcell model is considered as in~\cite{hoshino2011further}. We consider the bandwidth $B=20$ MHz. The coherence block contains $\tau_c =200$ channel uses, where $\tau_p=10$ channels are used for channel estimation. To evaluate the robustness of the proposed \ac{SUA} scheme, Monte Carlo simulations are conducted over multiple random network deployments. The performance of the proposed \ac{SUA} scheme is compared against the conventional unscalable \ac{CF-mMIMO} approach of \cite{7827017, 7917284, 8845768, 8476516}. The key performance metrics are mean \ac{SER}, probability of detection, position and velocity estimation, total energy consumption of \acp{AP}, transmission delay, and \ac{UA} runtime. Simulations are conducted using MATLAB version R2024b on a system equipped with an 8-core processor and 16 GB~RAM.
\subsection{Fine-Tuning \ac{SUA} Scheme Parameters}
\par
\begin{figure}
\centering 
\resizebox{0.8\columnwidth}{!}{
\includegraphics{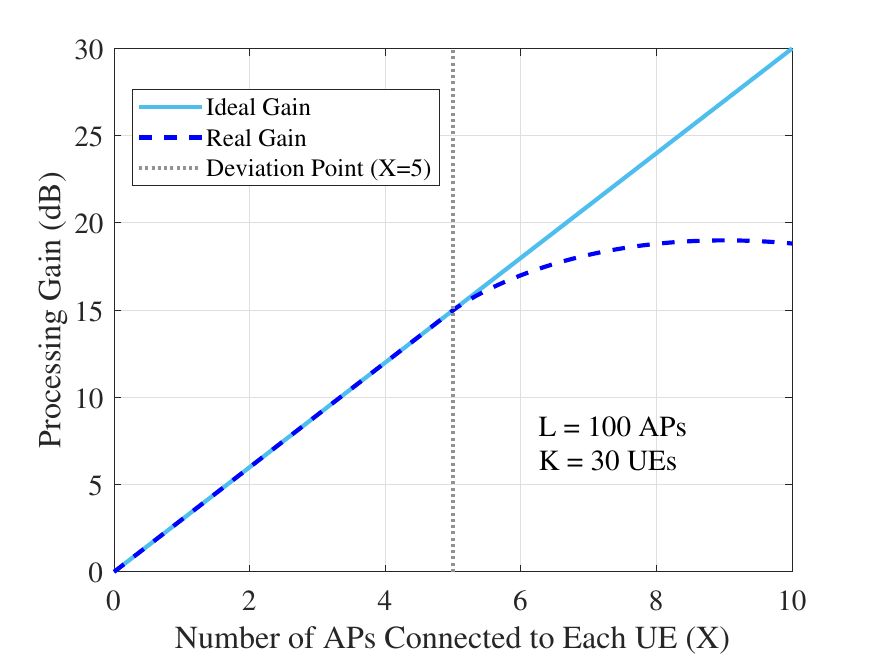}}
\caption{Ideal gain compared to the real gain with optimum value of $X$.}
\label{xx}
\end{figure}

 We next analyze the impact of varying \ac{SUA} parameters on network performance, focusing on the optimal number of \acp{AP} connected to each \ac{UE} i.e., $X$. Fig.~\ref{xx} shows the relationship between the maximum number of \acp{AP} connected to each \ac{UE} and the resulting system processing gain in decibels (dB). Processing gain here refers to the improvement in signal quality achieved by coherently combining signals from multiple \acp{AP}. The ideal gain 
represents the theoretical maximum gain, which increases linearly with $X$ under perfect conditions. The real gain 
shows the actual gain deviating from the ideal gain beyond $X=5$. This deviation indicates the threshold where additional \acp{AP} no longer yield proportional gain due to interference. Figure~\ref{xx} emphasizes the need to optimize $X$ for a balance between processing gain and efficiency.
Results show that $X=5$ is the optimal point to maximize efficiency under the considered conditions (i.e., $L=100$ \acp{AP} and $K=30$ \acp{UE}). Thus, following, we will use $X=5$. 
\par
\begin{figure}
\centering 
\resizebox{1\columnwidth}{!}{
\includegraphics{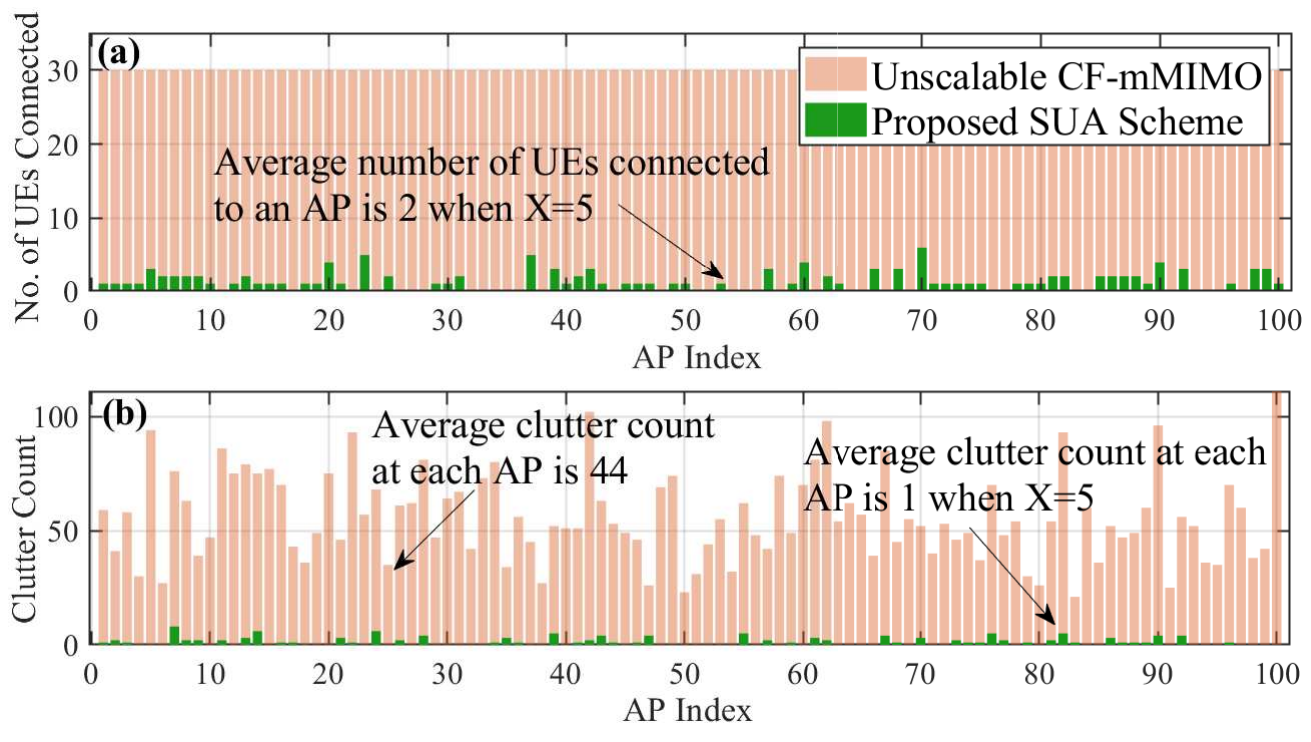}}
\caption{(a) Number of \acp{UE} served by each \ac{AP} index, and (b) clutter count affecting each \ac{AP} for unscalable \ac{CF-mMIMO} vs. the proposed \ac{SUA} scheme.} 
\label{double}
\end{figure}

 Fig. \ref{double} (a) compares the number of \acp{UE} connected to each \ac{AP} in the proposed \ac{SUA} scheme and unscalable \ac{CF-mMIMO} systems. For the unscalable approach, each \ac{AP} serves all 30 \acp{UE}, causing high front-haul traffic, computational complexity, and interference. In contrast, the proposed \ac{SUA} scheme  limits connections to a subset of \acp{AP} with $X=5$. As illustrated in Fig. \ref{xx}, this approach effectively balances scalability and performance, resulting in an average of two \acp{UE} connected to each \ac{AP}, with a maximum of six and a minimum of zero \acp{UE} per \ac{AP}. The occurrence of zero \acp{UE} at certain \acp{AP} indicates that the signal quality in those links is insufficient to establish a reliable connection, ensuring that only viable links are utilized for network efficiency. As we will see later, this reduces unnecessary connections, improving service~quality. 
\par Furthermore, Fig. \ref{double} (b) shows the number of clutter sources affecting each \ac{AP}. High clutter counts indicate interference from scatterers or other sources. In unscalable \ac{CF-mMIMO}, since all \acp{AP} are connected to all \acp{UE}, there is significant clutter interference for most \acp{AP}. This clutter degrades sensing and communication performance. In the unscalable \ac{CF-mMIMO} system, the average clutter count per link is approximately 44 scatterers, with a maximum of 111 and a minimum of 21 clutter scatterers. The proposed \ac{SUA} scheme reduces the clutter count significantly, by masking certain links and limiting \ac{AP}-\ac{UE} connections. In the proposed \ac{SUA} scheme, the average clutter count per link is approximately one scatterer, with a maximum of eight and a minimum of zero clutter scatterers. The optimization process eliminates weak and cluttered links. As a result, the clutter count per \ac{AP} is much lower, leading to improved system performance.
\begin{figure*}[ht]
\centering      
\subfloat[]{
  \includegraphics[width=58.04mm,height=46mm]{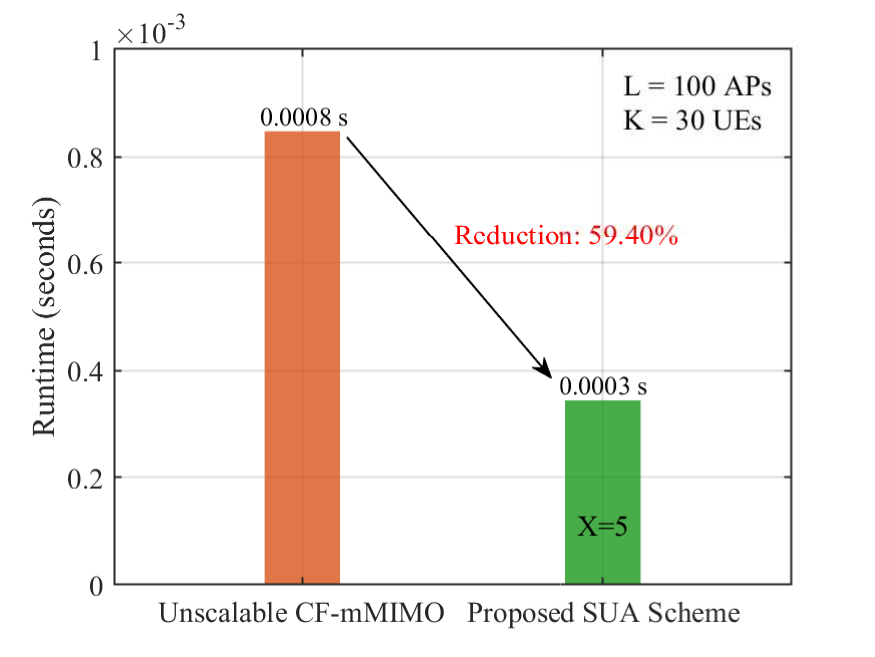}
}
\subfloat[]{
  \includegraphics[width=58.04mm,height=46mm] {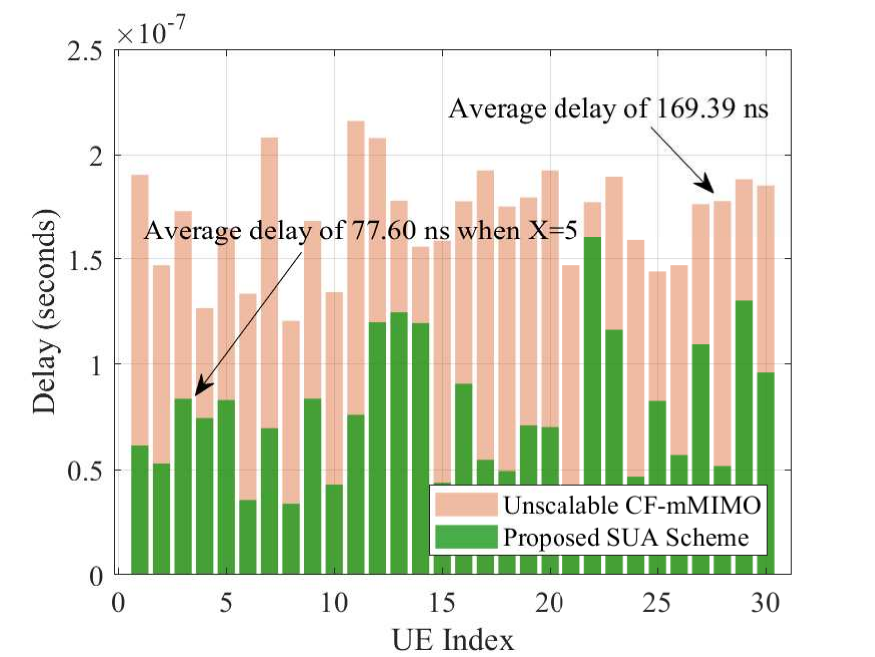}
}
\subfloat[]{
  \includegraphics[width=58.04mm,height=46mm]{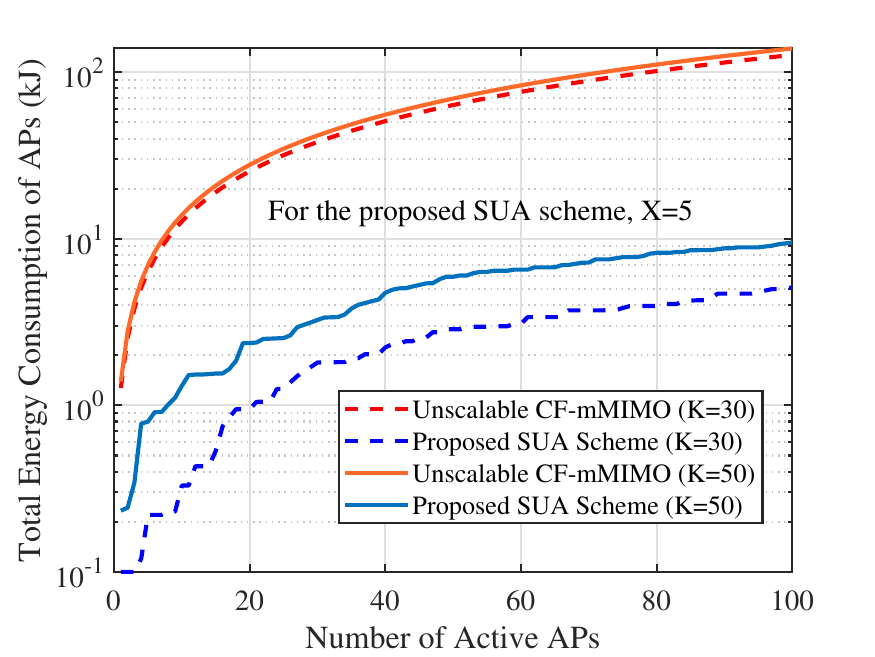}
}
\caption{(a) The \ac{UA} runtime comparison for unscalable \ac{CF-mMIMO} and the proposed \ac{SUA} scheme with $100$ \acp{AP} and $30$ \acp{UE}, (b) transmission delay experienced by each \ac{UE}, and (c) total energy consumption of the \acp{AP} as a function of active \acp{AP} for $30$ and $50$ \acp{UE} and $100$ \acp{AP}.}
\label{allBARS}
\end{figure*}
\subsection{\ac{CF-mMIMO} Network Performance}
\par This section analyzes various network-level performance aspects, including \ac{UA} runtime, transmission delay, and total energy consumption of \acp{AP}. These results highlight the significance of the proposed \ac{SUA} scheme, not just in improving individual link performance but in enhancing the overall network efficiency. Fig. \ref{allBARS} (a) shows the \ac{UA} runtime for the unscalable \ac{CF-mMIMO} systems and the proposed \ac{SUA} scheme. For the unscalable system, the runtime is approximately 0.0008 seconds, driven by resource over-allocation, excessive clutter, and suboptimal \ac{AP}-\ac{UE} associations. With the proposed \ac{SUA} scheme, the runtime is reduced to 0.0003 seconds, marking a 59.40$\%$ improvement. This reduction is achieved by masking weak links, prioritizing resources, and optimizing \ac{AP}-\ac{UE} associations, which streamlines processing and minimizes clutter-induced delays. The significant runtime reduction underscores the scalability of the proposed framework for real-time applications in dynamic, dense \ac{CF-mMIMO}~networks.
\par Fig. \ref{allBARS} (b) shows the transmission delay experienced by \acp{UE} for the unscalable \ac{CF-mMIMO} systems and the proposed \ac{SUA} scheme. The orange bars indicate significant delays for the unscalable \ac{CF-mMIMO} systems due to inefficient \ac{AP}-\ac{UE} associations, where distant \acp{AP} lead to high transmission delays. This affects communication and causes information aging, particularly for mobile \acp{UE}, as outdated data leads to inaccurate sensing. The average delay experienced by \acp{UE} in the unscalable \ac{CF-mMIMO} systems is around $196.39$ ns with a minimum and maximum delay of $120.642$ ns and $215.96$ ns, respectively. While for the proposed \ac{SUA} scheme, shown by the green bars, delays are significantly reduced due to improved \ac{AP}-\ac{UE} associations, addressing information aging and ensuring timely and accurate sensing and communication. For the proposed \ac{SUA} scheme, the average delay experienced by a \ac{UE} is around $77.60$ ns, while minimum and maximum delays experienced are $33.74$ ns and $160.63$ ns, respectively.  
\par Fig. \ref{allBARS} (c) shows the network's total energy consumption versus the varying number of active \acp{AP} for $K=30$ and $K=50$ \acp{UE}. The red curves represent the unscalable \ac{CF-mMIMO}, where all \acp{AP} connect to all \acp{UE}, leading to excessive energy use due to inefficient resource allocation. In contrast, the blue dashed and solid curves represent the proposed \ac{SUA} scheme, where only necessary \acp{AP} are activated, significantly reducing the energy consumption of the network while maintaining performance and highlighting the optimization's efficiency. As the number of users increases from $K=30$ to $K=50$ the total energy consumption of \acp{AP} rises for both the unscalable \ac{CF-mMIMO} and the proposed \ac{SUA} scheme. This increase occurs because more \acp{UE} require connectivity, leading to higher transmission power consumption and more active \acp{AP}. Moreover, the greater the number of \acp{UE}, the greater the difference, which means the proposed \ac{SUA} scheme is scalable.
\begin{figure}
\centering 
\resizebox{0.8\columnwidth}{!}{
\includegraphics{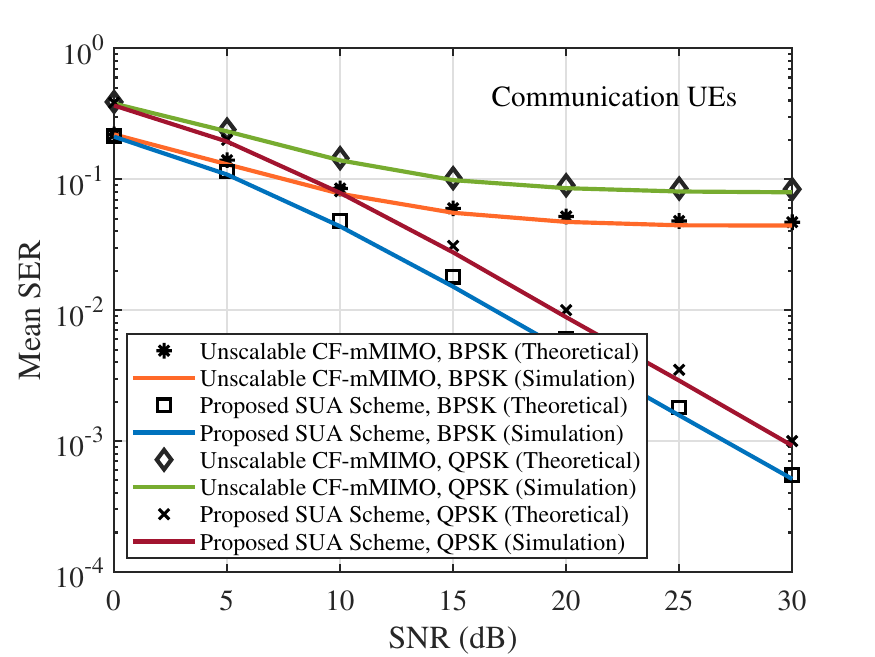}}
\caption{The mean \ac{SER} of communication \acp{UE} for the unscalable \ac{CF-mMIMO} and the proposed \ac{SUA} schemes.} 
\label{commuu}
\end{figure}
\subsection{Symbol Error Rate}
\par In this section we present the mean \ac{SER} for communication and \ac{JCAS} \acp{UE}. Fig. \ref{commuu} illustrates the mean \ac{SER} performance for both BPSK and QPSK modulations for communication \acp{UE}, comparing results for the unscalable \ac{CF-mMIMO} and the proposed \ac{SUA} scheme across varying \ac{SNR} values. The theoretical and simulated mean \ac{SER} values are presented for both scenarios, with the theoretical curves derived from Eq. (\ref{mainSERR}) serving as a benchmark for evaluating system performance. For the unscalable \ac{CF-mMIMO}, the mean \ac{SER} is significantly higher due to poor \ac{AP}-\ac{UE} associations, resulting in interference and suboptimal communication. While for the proposed \ac{SUA} scheme, the mean \ac{SER} improves drastically as effective \ac{AP}-\ac{UE} associations reduce interference and enhance signal quality. Notably, the theoretical and simulated results align closely, validating the efficacy of the proposed \ac{SUA} scheme and confirming the accuracy and practical applicability of the derived mathematical expressions. The improvement is particularly pronounced at higher \ac{SNR} levels, demonstrating that the optimization framework effectively supports robust communication under favorable conditions while maintaining consistent performance across modulation schemes.
\par
\begin{figure}[t!]
\centering 
\resizebox{0.8\columnwidth}{!}{
\includegraphics{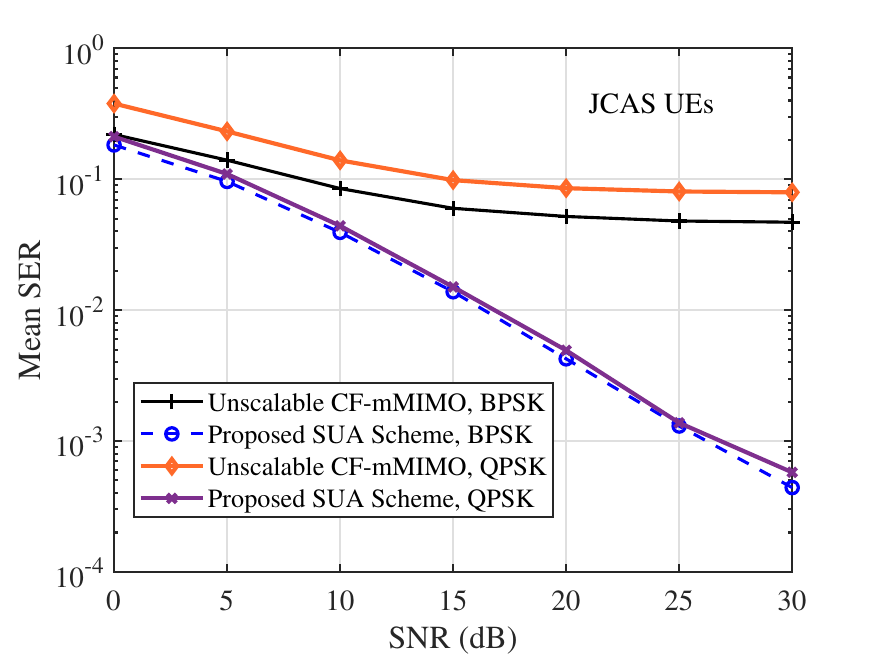}}
\caption{The mean \ac{SER} of \ac{JCAS} \acp{UE} for the unscalable \ac{CF-mMIMO} and the proposed \ac{SUA} schemes.} 
\label{commuu2}
\end{figure}

 Similar to Fig. \ref{commuu}, Fig. \ref{commuu2} also depicts the mean \ac{SER} for BPSK and QPSK modulated but for \ac{JCAS} \acp{UE} as a function of \ac{SNR} for the unscalable \ac{CF-mMIMO} and the proposed \ac{SUA} scheme. The black and orange curves represent the mean \ac{SER} for the unscalable \ac{CF-mMIMO}, which is consistently higher across all \ac{SNR} values due to high interference from neighboring links. In contrast, the blue and purple curves indicate the proposed \ac{SUA} scheme, showing a significant reduction in \ac{SER} as the optimization framework eliminates inefficient \ac{AP}-\ac{UE} links and prioritizes high-quality connections. The sharp decrease in \ac{SER}, particularly at lower \ac{SNR} values, highlights the robustness of the proposed \ac{SUA} scheme. Moreover, mean \ac{SER} of communication \acp{UE} is better compared to that of \ac{JCAS} \acp{UE} because of the duality functionality in \ac{JCAS} compared to that for only communication \acp{UE}.
\begin{figure}[h!]
\centering  
\subfloat[]{
  \includegraphics[width=41.91mm,height=38mm]{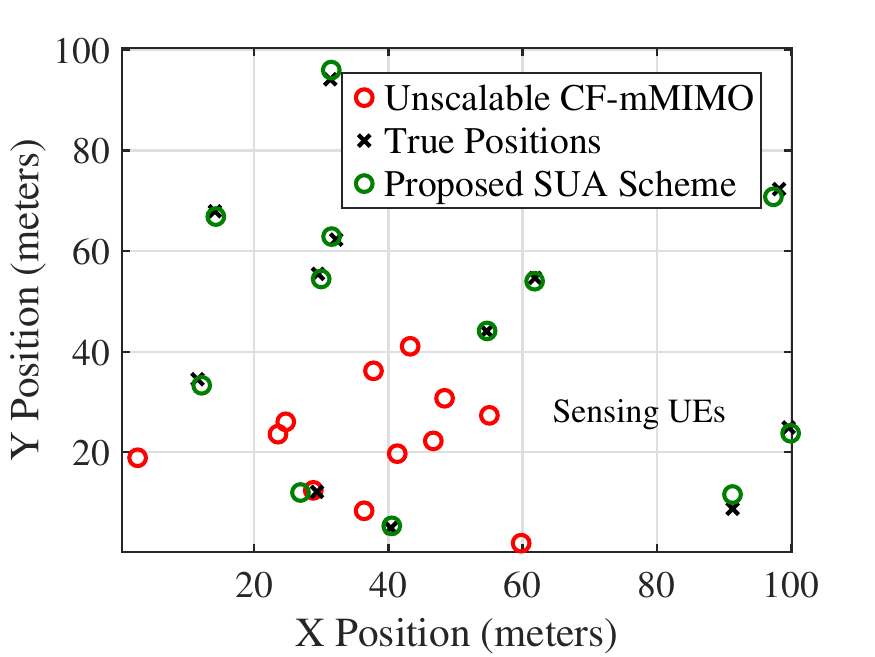}
}
\subfloat[]{
  \includegraphics[width=41.91mm,height=38mm]{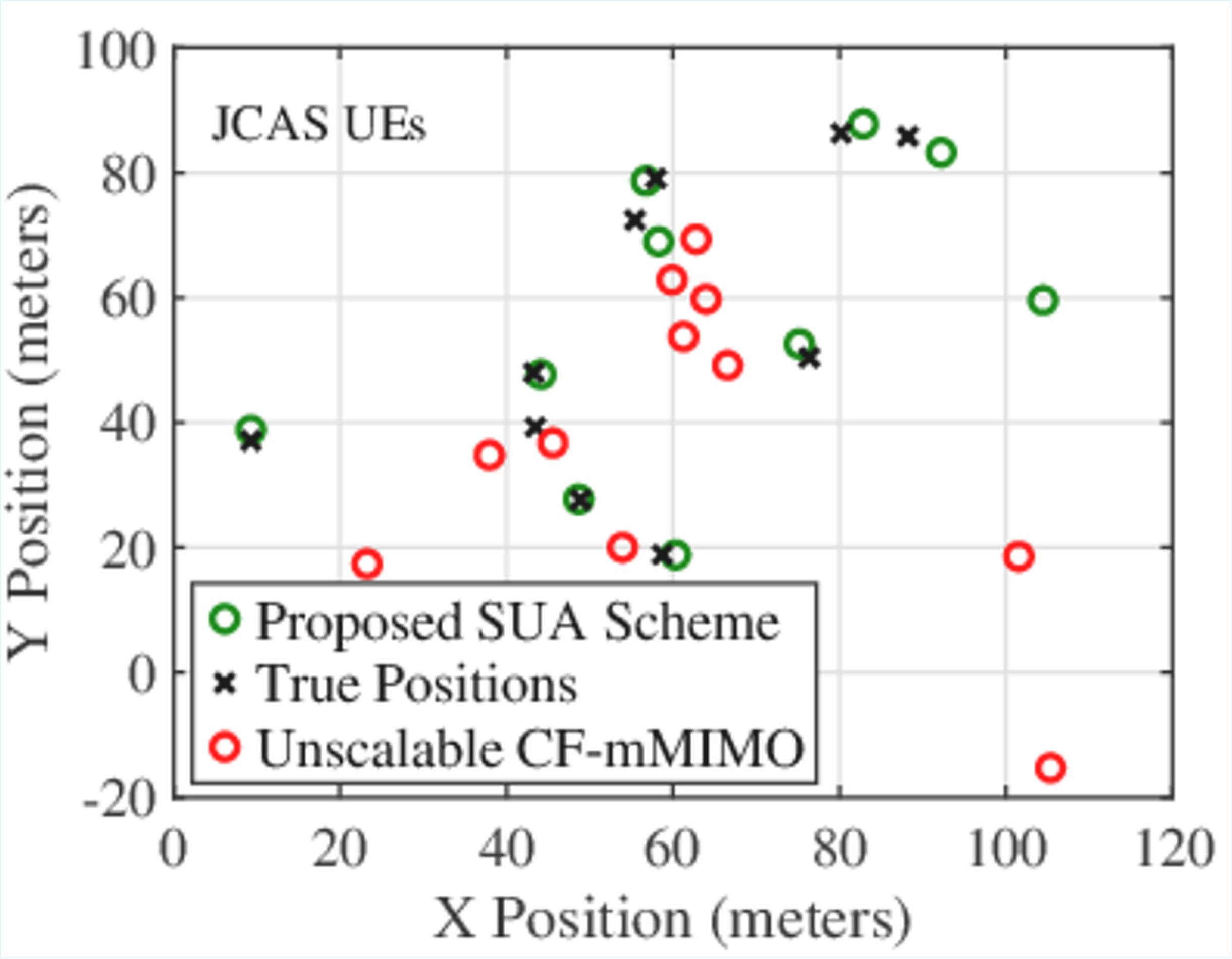}
}
\hspace{0mm}
\subfloat[]{
  \includegraphics[width=41.91mm,height=38mm]{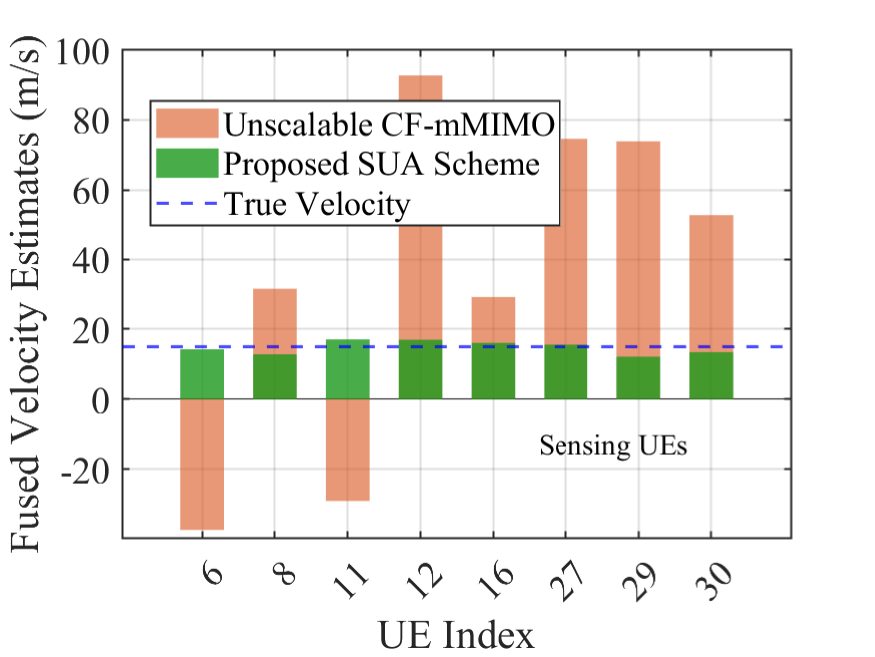}
}
\subfloat[]{
  \includegraphics[width=41.91mm,height=38mm]{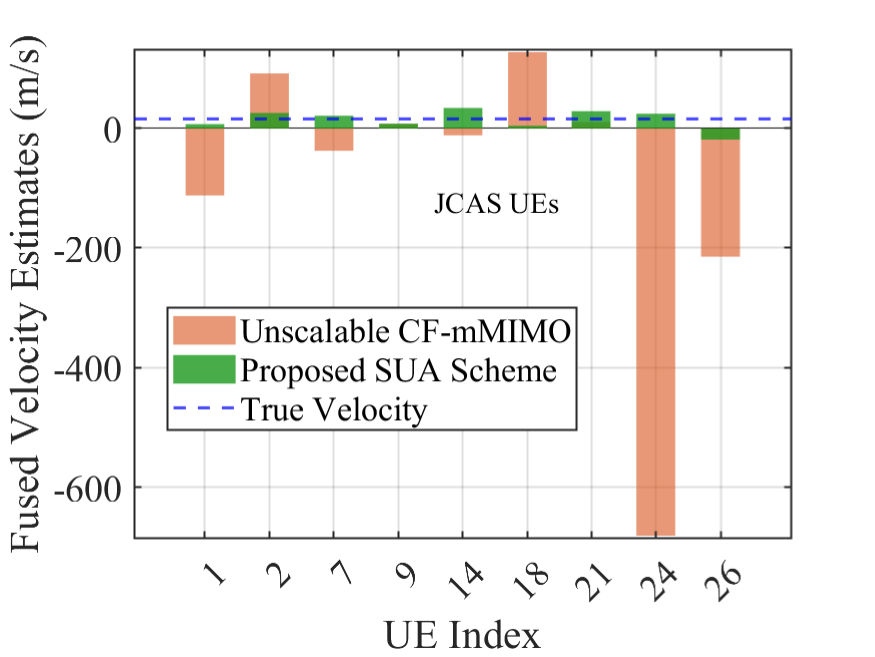}
}
\caption{The UE position estimates for the unscalable \ac{CF-mMIMO} and the proposed \ac{SUA} schemes for (a) sensing \acp{UE} and (b) \ac{JCAS} \acp{UE}, comparison of fused velocity estimates for the unscalable \ac{CF-mMIMO} and the proposed \ac{SUA} schemes for (c) sensing \acp{UE} and (d) \ac{JCAS} \acp{UE}.}
\label{positions}
\end{figure}
\subsection{Position Estimation of the \acp{UE}}
\par This section details the process by which each \ac{AP} transmits its sensed data to the \ac{CPU}, where data fusion techniques are applied to accurately estimate the precise location of each \ac{UE}. Fig. \ref{positions} compares the position estimation accuracy of \acp{UE} for the unscalable \ac{CF-mMIMO} and the proposed \ac{SUA} scheme for sensing and \ac{JCAS} \acp{UE}. In the figure, red circles represent the estimated positions for the unscalable \ac{CF-mMIMO}, green circles represent positions for the proposed \ac{SUA} scheme, and black crosses indicate the true \ac{UE} positions. For sensing \acp{UE} in Fig. \ref{positions} (a), the estimates for the unscalable \ac{CF-mMIMO} show significant deviation from the true positions due to clutter interference and weak \ac{AP}-\ac{UE} connectivity. While for the proposed \ac{SUA} scheme, the green circles closely align with the true positions, demonstrating improved accuracy.
\begin{figure*}[h]
\centering  
\subfloat[]{
  \includegraphics[width=58.04mm,height=45mm]{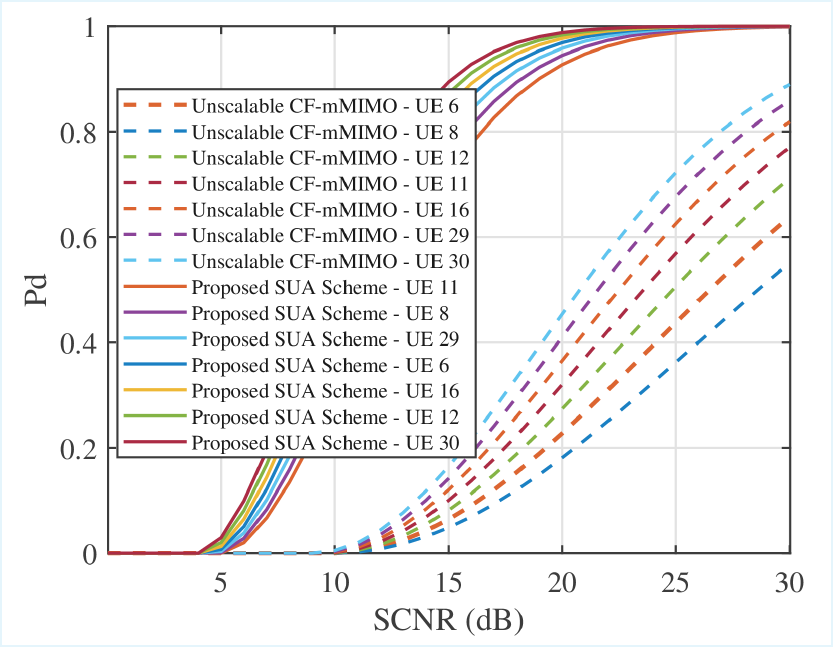} 
}
\subfloat[]{
  \includegraphics[width=58.04mm,height=45mm]{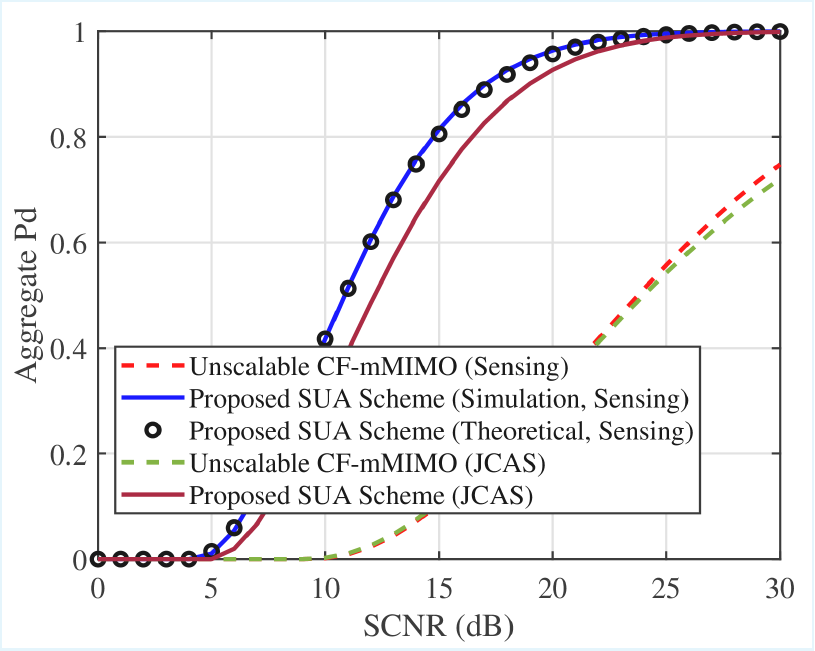}
}
\subfloat[]{
  \includegraphics[width=58.04mm,height=45mm] {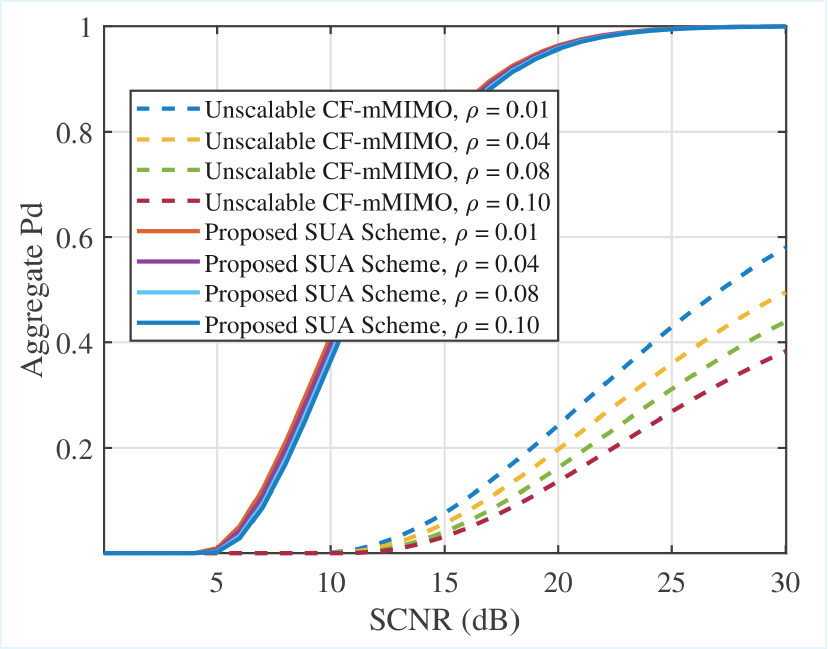}
} 
\caption{(a) \ac{$P_d$} for individual sensing \acp{UE} as a function of \ac{SCNR}, (b) aggregate \ac{$P_d$} as a function of \ac{SCNR} for sensing and \ac{JCAS} \acp{UE}, and (c) \ac{$P_d$} vs. \ac{SCNR} for different clutter densities for the unscalable \ac{CF-mMIMO} and the proposed \ac{SUA} schemes.}
\label{thing2}
\end{figure*}
\par Similarly, Fig. \ref{positions} (b) presents the position estimates for \ac{JCAS} users for the unscalable \ac{CF-mMIMO} and the proposed \ac{SUA} scheme. While \ac{JCAS} position estimates exhibit slightly lower accuracy than those of sensing \acp{UE}, due to the dual task of sensing and communication, the optimization significantly enhances performance. By reducing clutter and ensuring more reliable measurements, the optimized scheme achieves notable improvements in position estimation accuracy for \ac{JCAS} \acp{UE}.
\subsection{Velocity Estimation of the \acp{UE}}
\par In this section we analyze the velocity estimates for sensing and \ac{JCAS} \acp{UE}. Fig. \ref{positions} (c) shows the fused velocity estimates of sensing \acp{UE} for the unscalable \ac{CF-mMIMO} and the proposed \ac{SUA} scheme, compared to the true velocity of 15 m/s. For the unscalable \ac{CF-mMIMO} (orange bars), velocity estimates deviate significantly from the true value due to high clutter, and weak radar echo returns. The average velocity for the unscalable \ac{CF-mMIMO} system is $52.64$ m/s, with a maximum estimated velocity of $92.67$~m/s and a minimum of $29.12$ m/s. While the proposed \ac{SUA} scheme, represented by the green bars, the estimates align closely with the true velocity, highlighting improved accuracy. The optimization reduces errors by selecting links with the least clutter which enhances the accuracy of velocity estimation. For the proposed \ac{SUA} scheme, the average, maximum, and minimum velocities are  $14.77$ m/s, $17.07$ m/s, and $12.10$ m/s respectively.
\par Moreover, the chart in Fig. \ref{positions} (d) shows the fused velocity estimates for \ac{JCAS} \acp{UE} compared to the true velocity (i.e., 15 m/s). For the unscalable \ac{CF-mMIMO}, the velocity estimates deviate drastically, with some values exceeding $-400$ m/s or $100$ m/s, due to excessive clutter and suboptimal \ac{UA}. The average velocity estimate of $144.71$ m/s, a maximum estimate of $680.82$ m/s, and a minimum estimate of $11.41$ m/s. While in the proposed \ac{SUA} scheme, the velocity estimates are much closer to the true value, indicating improved accuracy. For the proposed \ac{SUA} scheme, an average velocity estimate  $19.15$ m/s, a maximum of $34.23$ m/s, and a minimum of $7.12$ m/s is experienced. While \ac{JCAS} sensing performance is slightly degraded compared to sensing \acp{UE} due to the dual objectives of communication and sensing, it still significantly improves after optimization. This is because the optimization process reduces clutter and ensures that echoes received by \acp{AP} primarily originate from the \acp{UE} rather than unwanted reflections, thus enhancing sensing accuracy even in \ac{JCAS}~setup. 
\subsection{Probability of Detection}
\par This section relates the \ac{$P_d$} as a sensing performance metric and highlights the proposed optimization's impact on \ac{$P_d$} under various scenarios. Fig. \ref{thing2} (a) shows \ac{$P_d$} for individual sensing \acp{UE} as a function of \ac{SCNR}. For the unscalable \ac{CF-mMIMO} (dashed line), \ac{$P_d$} is lower and varies widely due to high clutter. While for the proposed \ac{SUA} scheme (solid line), \ac{$P_d$} is higher and more consistent, reflecting improved sensing performance through reduced clutter and better AP-UE associations. 
\par Fig. \ref{thing2} (b) presents aggregate \ac{$P_d$} across the sensing \acp{UE} and \ac{JCAS} \acp{UE}. For the sensing \acp{UE} in unscalable \ac{CF-mMIMO} (red dashed line), aggregate \ac{$P_d$} is lower, especially at low \ac{SCNR}, due to clutter and inefficient resource allocation. While for the proposed \ac{SUA} scheme (blue solid line), \ac{$P_d$} significantly improves and aligns closely with the theoretical model, validating the optimization framework's accuracy. Moreover, Fig. \ref{thing2} (b) also shows the aggregate \ac{$P_d$} for the \ac{JCAS} \acp{UE}. The unscalable \ac{CF-mMIMO} (green dashed line) has a lower \ac{$P_d$} compared to that of the proposed \ac{SUA} scheme (maroon solid line). Whereas \ac{JCAS} \acp{UE} has to balance sensing and communication, resulting in marginally lower \ac{$P_d$} compared to the sensing \acp{UE}. Both optimized cases demonstrate marked improvement, highlighting the optimization scheme's effectiveness in enhancing detection while accommodating diverse \ac{UE} requirements. Fig. \ref{thing2} (c) examines the effect of clutter density on aggregate \ac{$P_d$}. For the unscalable \ac{CF-mMIMO} (dashed lines), \ac{$P_d$} decreases with higher clutter density. While for the proposed \ac{SUA} scheme (solid lines), \ac{$P_d$} remains stable across clutter densities and the nearly overlapping curves demonstrate that the optimization effectively mitigates the impact of clutter, ensuring consistent sensing performance.
\section{Conclusion}
This paper proposed a \ac{SUA} scheme for \ac{CF-mMIMO}, addressing the two main challenges of serving heterogeneous \ac{UE} requirements and having a scalable \ac{UA}. By integrating AP masking, link prioritization, and optimization-based \ac{UA}, the proposed SUA scheme effectively selects the most relevant \acp{AP} for each \ac{UE} based on communication, sensing, and \ac{JCAS} requirements. Simulation results demonstrated significant improvements in mean \ac{SER} and \ac{$P_d$}, while reducing APs' energy consumption and computational runtime. The results also highlighted the importance of fine-tuning the \ac{AP}-\ac{UE} associations to balance performance and scalability. Future work will explore dynamic user mobility and machine learning-based optimization techniques to further enhance network efficiency and adaptability in real-time network scenarios.
{\appendix [Proof of Theorem 1]}
The output of the \ac{ML} detector can be denoted as, $\hat{\mathbf{s}}_{\mathrm{ML}} = \arg \min_{s{_k} \in \mathbb{S}} \| \mathbf{y} - \hat{\mathbf{h}}_{lk} s_k \|^2,$ where $\mathbb{S}$ is the set of all possible symbols. The \ac{PEP} quantifies the probability that transmitted symbol $s_k^i$ is erroneously detected as $s_k^j$, $\Pr({s}_k^i \to {s}_k^j | \mathbf{\hat{h}}_{lk})$. For a \ac{ML} detector, the decision is based on minimizing the Euclidean distance between the received signal \(\mathbf{y}\) and the hypothesis \(\mathbf{\hat{h}}_{lk}s\). The event \({s}_k^i \to {s}_k^j\) occurs if $\|\mathbf{y} - \mathbf{\hat{h}}_{lk}{s}_k^i\|^2 \geq \|\mathbf{y} - \mathbf{\hat{h}}_{lk}{s}_k^j\|^2.$ So, the upper bound of the \ac{cPEP} is:
\begin{equation} \label{p1}
\Pr({s}_k^i \to {s}_k^j | \mathbf{\hat{h}}_{lk})=\Pr (\|\mathbf{y} - \mathbf{\hat{h}}_{lk}{s}_k^i\|^2 \geq \|\mathbf{y} - \mathbf{\hat{h}}_{lk}{s}_k^j\|^2).
\end{equation}
The received signal $\mathbf{y}$ is expressed as $\mathbf{y}=\mathbf{h}_{lk}{s}_k^i+\mathbf{n}$, $\mathbf{y} = \mathbf{\hat{h}}_{lk}{s}_k^i + \mathbf{\tilde{h}}_{lk}{s}_k^i + \mathbf{n}$, where $\mathbf{\tilde{h}}_{lk}$ is the channel estimation error. By substituting $\mathbf{y}$ in the inequality (\ref{p1}), we get:
\begin{equation} \label{p_combined3}
\begin{aligned}
&\Pr({s}_k^i \to {s}_k^j | \mathbf{\hat{h}}_{lk}) = 
\Pr\left(\|\big(\mathbf{\hat{h}}_{lk}{s}_k^i + \mathbf{\tilde{h}}_{lk}{s}_k^i + \mathbf{n}\big) 
- \mathbf{\hat{h}}_{lk}{s}_k^i\|^2 \geq \right. \\& \left. \|\big(\mathbf{\hat{h}}_{lk}{s}_k^i + \mathbf{\tilde{h}}_{lk}{s}_k^i + \mathbf{n}\big) - \mathbf{\hat{h}}_{lk}{s}_k^j\|^2\right) = \Pr\left(\|\mathbf{\tilde{h}}_{lk}{s}_k^i + \mathbf{n}\|^2\geq \right. \\& \left. \|\mathbf{\tilde{h}}_{lk}{s}_k^i + \mathbf{n} + \right. \left. \mathbf{\hat{h}}_{lk}({s}_k^i - {s}_k^j)\|^2\right).
\end{aligned}
\end{equation}
Furthermore, by using the sum square property of Euclidean norm expansion as $\|\mathbf{a} + \mathbf{b}\|^2 = \|\mathbf{a}\|^2 + \|\mathbf{b}\|^2 + 2\text{Re}(\mathbf{a}^H\mathbf{b})$, we expand (\ref{p_combined3}). The left-hand side of (\ref{p_combined3}) becomes $\|\mathbf{\tilde{h}}_{lk}{s}_k^i + \mathbf{n}\|^2 = \|\mathbf{\tilde{h}}_{lk}{s}_k^i\|^2 + \|\mathbf{n}\|^2 + 2\text{Re}((\mathbf{\tilde{h}}_{lk}{s}_k^i)^H\mathbf{n})$, and the right-hand side is as $\|\mathbf{\tilde{h}}_{lk}{s}^i + \mathbf{n} + \mathbf{\hat{h}}_{lk}({s}^i - {s}^j)\|^2 = \|\mathbf{\tilde{h}}_{lk}{s}^i + \mathbf{n}\|^2 + \|\mathbf{\hat{h}}_{lk}({s}^i - {s}^j)\|^2 + 2\text{Re}((\mathbf{\tilde{h}}_{lk}{s}^i + \mathbf{n})^H\mathbf{\hat{h}}_{lk}({s}^i - {s}^j))$. By canceling common terms \(\|\mathbf{\tilde{h}}_{lk}{s}_k^i + \mathbf{n}\|^2\) on both sides:
{\small
\begin{equation} \label{p4}
\begin{aligned}
&\Pr \left(0 \geq \|\mathbf{\hat{h}}_{lk}({s}_k^i - {s}_k^j)\|^2 + 2\text{Re}((\mathbf{\tilde{h}}_{lk}{s}_k^i + \mathbf{n})^H\mathbf{\hat{h}}_{lk}. ({s}_k^i - {s}_k^j))\right)\\&=\Pr \left(2\text{Re}((\mathbf{\tilde{h}}_{lk}{s}_k^i + \mathbf{n})^H\mathbf{\hat{h}}_{lk}({s}_k^i - {s}_k^j)) \leq-\|\mathbf{\hat{h}}_{lk}({s}_k^i - {s}_k^j)\|^2 \right).
\end{aligned}
\end{equation}}
\par To analyze the behavior of the decision metric under the effects of \ac{CEE} and noise, we define the decision metric $J$ as, $J = \left[\mathbf{\hat{h}}_{lk}(s_k^i - s_k^j)\right]^H \left(\mathbf{n} + \mathbf{\tilde{h}}_{lk}s_k^j\right)$, where expectation $\mathbb{E}(J)$, is derived with respect to $\mathbf{n}$ and $\mathbf{\tilde{h}}_{lk}$. Using linearity of $\mathbb{E}(J)$ and the independence of $\mathbf{n}$ and $\mathbf{\tilde{h}}_{lk}$ we have, $\mathbb{E}(J) = \left[\mathbf{\hat{h}}_{lk}(s_k^i - s_k^j)\right]^H \mathbb{E}\left\{\mathbf{n} + \mathbf{\tilde{h}}_{lk}s_k^j\right\}$. Since $\mathbf{n}$ and $\mathbf{\tilde{h}}_{lk}$ are zero-mean random variables, $\mathbb{E}(J) = \left[\mathbf{\hat{h}}_{lk}(s_k^i - s_k^j)\right]^H \mathbb{E}\left\{\mathbf{n} + \mathbf{\tilde{h}}_{lk}s_k^j\right\}=0$, with variance of $\mathbb{D}(J)$:
\begin{equation} \label{p9}
\mathbb{D}(J) = \left[\mathbf{\hat{h}}_{lk}(s_k^i - s_k^j)\right]^H \mathbb{D}\left\{\mathbf{n} + \mathbf{\tilde{h}}_{lk}s_k^j\right\} \left[\mathbf{\hat{h}}_{lk}(s_k^i - s_k^j)\right],
\end{equation}
where the variance of $\mathbf{n} + \mathbf{\tilde{h}}_{lk}s_k^j$ is $\mathbb{D}\{\mathbf{n}\} + \mathbb{D}\{\mathbf{\tilde{h}}_{lk}s^j\}$. Since \(\mathbf{n}\) is white Gaussian noise with variance $\sigma^2$ then $\mathbb{D}\{\mathbf{n}\} = \sigma^2 \mathbf{I}_{N}$,  where $\mathbf{I}_{N}$ is the identity matrix. Moreover, $\mathbf{\tilde{h}}_{lk}$ and $s_k^j$ are independent, so $\mathbb{D}\{\mathbf{\tilde{h}}_{lk}s_k^j\} = \mathbb{E}\{\mathbf{\tilde{h}}_{lk}s_k^j(\mathbf{\tilde{h}}_{lk}s_k^j)^H\}$. Let $\mathbf{B}_k$ denote the covariance matrix of the channel estimation error, then $\mathbb{D}\{\mathbf{\tilde{h}}_{lk}s_k^j\} = \sum_{k=1}^{K} p_k \mathbf{B}_k$. Combining these results:
\begin{equation} \label{p11}
\mathbb{D}\{\mathbf{n} + \mathbf{\tilde{h}}_{lk}s_k^j\} = \sigma^2 \mathbf{I}_{N} + \sum_{k=1}^{K} p_k \mathbf{B}_k.
\end{equation}
Substituting Eq. (\ref{p11}) into the expression for $\mathbb{D}(J)$ in Eq. (\ref{p9}):
{\small
\begin{equation} 
\begin{aligned}
\mathbb{D}(J) =& \left[\mathbf{\hat{h}}_{lk}(s_k^i - s_k^j)\right]^H \left(\sigma^2 \mathbf{I}_{N} + \sum_{k=1}^{K} p_k \mathbf{B}_k\right) \left[\mathbf{\hat{h}}_{lk}(s_k^i - s_k^j)\right], \\
 =& \left[\mathbf{\hat{h}}_{lk}(s_k^i - s_k^j)\right]^H \sigma^2 \mathbf{I}_{N} \left[\mathbf{\hat{h}}_{lk}(s_k^i - s_k^j)\right] 
+ \left[\mathbf{\hat{h}}_{lk}(s_k^i - \right. \\& \left. s_k^j)\right]^H \left(\sum_{k=1}^{K} p_k \mathbf{B}_k\right) \left[\mathbf{\hat{h}}_{lk}(s_k^i - s_k^j)\right]. \\
\end{aligned}
\end{equation}}
Using Gaussian distribution properties and the relationship with the $Q(x) = \int_{x}^{\infty} \frac{1}{\sqrt{2\pi}} e^{-t^2/2} \, dt$, we express the \ac{cPEP} as:

{\small
\begin{equation} \label{p_combined2}
\begin{aligned}
 \Pr&\left(s_k^i \to s_k^j \big|  \mathbf{\hat{h}}_{lk}\right) 
 = Q\left( \frac{\|\mathbf{\hat{h}}_{lk}(s_k^i - s_k^j)\|^2 / 2 - \mathbb{E}(J)}{\sqrt{\mathbb{D}(J)/2}} \right), \\
& = Q\left( 
\frac{\|\mathbf{\hat{h}}_{lk}(s_k^i - s_k^j)\|^2 / 2}
{\sqrt{\frac{1}{2} \left[\mathbf{\hat{h}}_{lk}(s_k^i - s_k^j)\right]^H \Sigma \left[\mathbf{\hat{h}}_{lk}(s_k^i - s_k^j)\right]}}
\right), \\
& = Q\left(
\frac{\|\mathbf{\hat{h}}_{lk}(s_k^i - s_k^j)\|^2}
{\sqrt{2 \left[\mathbf{\hat{h}}_{lk}(s_k^i - s_k^j)\right]^H \Sigma \left[\mathbf{\hat{h}}_{lk}(s_k^i - s_k^j)\right]}}
\right).
\end{aligned}
\end{equation}
}


\end{document}